\documentclass[12pt]{article}
\usepackage{epsf}
\title{Valence quark distributions in mesons \\
in generalized QCD sum rules.}
\author{B.L.Ioffe\thanks{Email address: ioffe@vitep5.itep.ru} and
A.G.Oganesian\thanks{Email address: armen@vitep5.itep.ru}\\
Institute of Theoretical and Experimental Physics,\\
B.Cheremushkinskaya 25, 117218 Moscow,Russia}
\date{}
\begin{document}
\maketitle

\newcommand{\be}{\begin{equation}}
\newcommand{\ee}{\end{equation}}

\def\la{\mathrel{\mathpalette\fun <}}
\def\ga{\mathrel{\mathpalette\fun >}}
\def\fun#1#2{\lower3.6pt\vbox{\baselineskip0pt\lineskip.9pt
\ialign{$\mathsurround=0pt#1\hfil##\hfil$\crcr#2\crcr\sim\crcr}}}

\begin{abstract}
The method for calculation of valence quark distributions at intermediate
$x$ is presented. The imaginary part of the virtual photon forward
scattering amplitude on the quark current with meson quantum number is
considered. Initial and final virtualities $p^2_1$ and $p^2_2$ of the
currents are assumed to be large, negative and different, $p^2_1 \not=
p^2_2$. The operator product expansion (OPE) in $p^2_1, p^2_2$  up to
dimension 6 operators is performed. Double dispersion representations in
$p^2_1, p^2_2$ of the amplitude in terms of physical states contributions
are used. Equalling them to those calculated in QCD by OPE the desired sum
rules for quark distributions in mesons are found. The double Borel
transformations are applied to the sum rules, killing non-diagonal
transition terms, which deteriorated the accuracy in the previous
calculations of quark distributions in nucleon. Leading order perturbative
corrections are accounted. Valence quark distributions in pion,
longitudinally and transversally polarized $\rho$-mesons are calculated at
intermediate $x$, $0.2 \la x \la 0.7$ and normalization points $Q^2 = 2-4
~GeV^2$ with no fitting parameters. The use of the Regge behaviour at small
$x$ and quark counting rules at large $x$ allows one to find the first and
the second moments of valence quark distributions. The obtained quark
distributions may be used as an input for evolution equations. In the case
of pion the quark distribution is in agreement with those found from the
data on the Drell-Yan process.
Quark distributions in transversally and longitudinally polarized
$\rho$-mesons are essentially different.

\end{abstract}

\newpage

\section{Introduction}

Quark and gluon distributions in hadrons are not calculated in QCD starting
from the first principles. What is only possible is to calculate their
evolution with $Q^2$. In the case of nucleon the standard way is the
following. At some fixed $Q^2 = Q^2_0$ the form of quark and gluon
distributions, characterized by the number of free parameters is assumed.
Then the evolution of distributions with $Q^2$ is calculated and after
comparing with the data of the deep inelastic lepton-nucleon scattering
the best fit for these parameters is found. (About dozen of the parameters
is used,  see, e.g. the recent papers \cite{1}-\cite{4}). Quark and gluon
distributions in pions are determined in a similar way from the data on
production of $e^+e^-$ and $\mu^+ \mu^-$ pairs in pion-nucleon collisions
(the Drell-Yan process) \cite{5,6}
 and prompt photons production \cite{ab}.  But here
one needs an additional hypothesis  about connection of the fragmentation
function at time-like $Q^2$, which is measured in the Drell-Yan processes,
with parton distributions defined at space-like $Q^2$.  For other mesons and
baryons, as well as for unmeasured up to now distributions in polarized
nucleon (like $h_1(x)$) we have no information about parton distributions
from experiment and have to rely on models. The QCD calculation of quark and
gluon distributions in hadrons, which could be used as an input in evolution
equations may be considered as a challenge for the theory.

The method of valence quark distributions calculation at intermediate $x$
was suggested in \cite{7} and developed in \cite{8}-\cite{10}. The idea was
to consider the imaginary part (in $s$-channel) of 4-point correlator,
corresponding to the forward scattering of two-quark currents, one of which
has the quantum numbers of hadron of interest and the other is
electromagnetic (or weak).  It was supposed that virtualities of the photon
and hadronic current $q^2$ and $p^2$ are large and negative $\vert q^2 \vert
\gg \vert p^2 \vert \gg R^{-2}_c$ , where $R_c$ is the confinement radius
($q$ is the momentum of virtual photon, $p$ is the momentum, carried by
hadronic current). It was shown \cite{8}, that in this case the imaginary
part in $s$ channel [$s = (p +q)^2$] of the forward scattering amplitude is
dominated by small distance contributions at intermediate $x$. (The standard
notation is used: $x$ is the Bjorken scaling variable , $x = -q^2/2 \nu, ~
\nu = pq$). The proof of this statement, given in \cite{8}, is based on the
fact that for the imaginary part of the forward scattering amplitude the
position of the closest to zero singularity in momentum transfer is
determined by the boundary of the Mandelstam spectral function and is given
by the equation

\be
t_0 = -4 \frac{x}{1-x} p^2
\label{1}
\ee
Therefore, if $\vert p^2 \vert$ is large and $x$ is not small, then even at
$t = 0$ (the forward amplitude) the virtualities of intermediate states in
$t$-channel are large enough for OPE to be applicable. The further procedure
is common for the QCD sum rules. On one side, the amplitude is calculated
by OPE with the account of condensates. On the other side, the dispersion
representation in $p^2$ in terms of physical states is written for the same
amplitude and the contribution of the lowest state is extracted by using the
Borel transformation. As follows from (\ref{1}), the approach is invalid at
small $x$. (Since in real calculations $-p^2$ are of order $1-2~ GeV^2$,
small $x$ means in fact $x \la 0.1 - 0.2$). This statement is evident
{\it apriori}, because at small $x$ Regge behaviour is expected, which
cannot be described in the framework of OPE. The approach is also invalid at
$x$ close to 1. This is the domain of resonances, also outside the scope of
OPE.  The fact that this method of calculation of quark distributions in
hadrons is invalid at $x \ll 1$ and at $1 - x \ll 1$ follows from the theory
itself:  the OPE diverges in these two domains. Therefore, calculating
higher order terms of OPE makes it possible to estimate up to which
numerical values of $x$, in the small and large $x$ domain, the theory is
reliable in each particular case. In the way described above, valence quark
distributions in nucleon \cite{8,10} were calculated. However, the accuracy
of the calculation was not good enough, especially for $d$-quarks \cite{8}.
Moreover, it was found to be impossible to calculate quark distributions in
$\pi$- and $\rho$-mesons in this way. The reason is that the sum rules in
the form used in \cite{8,10} have a serious drawback. (The calculation of
the photon structure function \cite{9} is a special case and has no such
problem).

The origin of this drawback comes from the fact that in the case of 4-point
function, which corresponds to the forward scattering amplitude with equal
inital and final hadron momenta, the Borel transformation does not provide
suppression of all excited state contributions: the non-diagonal matrix
elements, like

\be
\langle 0 \vert j^h \vert h^*\rangle\langle h^* \vert j^{el} (x) j^{el} (0)
h \rangle \langle h \vert j^h \vert 0 \rangle \label{2}
\ee
are not
suppressed in comparison with the matrix element of interest

\be
\langle 0 \vert j^h \vert h \rangle \langle h \vert j^{el} (x) j^{el} (0)
\vert h \rangle \langle h \vert j^h \vert 0 \rangle
\label{3}
\ee
proportional to the hadron $h$ structure function. (Here $h$ is the hadron
which structure function we would like to calculate, $h^*$ is the
excited state with the same quantum numbers as $h$, $j^h$ is the quark
current with quantum numbers of hadron $h$, $j^{el}$ is the electromagnetic
current). In order to kill the background matrix elements (\ref{2}) it was
necessary to differentiate the sum rule over the Borel parameter. But, as
is well known, differentiation of an approximate relation may seriously
deteriorate the accuracy of the results. In QCD sum rules such
differentiation increases contributions of higher order terms of OPE and
excited states in physical spectrum, the sum rules become much worse or even
fail (as for $\pi$ and $\rho$-mesons). For pion the situation is especially
bad, because direct calculation show, that the leading term in OPE (the bare
loop diagram) corresponds just to the non-diagonal matrix element, not to
the pion structure function.

In ref.12 it was suggested the modified method of calculation of the
hadron structure functions (quark distributions in hadrons), where this
problem is eliminated and valence quark distributions in pion were
calculated. This method is used here for calculation of valence quark
distributions in $\rho$-meson, separately for longitudinally and
transversally (relative to the virtual photon beam) polarized $\rho$-mesons.
Since quark distributions in  $\rho$-meson cannot be measured
experimentally, the common way to get them is to assume SU(6) symmetry,
where $\pi$- and $\rho$-mesons belong to the same multiplet. Then quark
distributions in $\rho$-meson are equal to that in pion and are independent
on $\rho$ polarization.
On the other side, the pion play a specific role in the theory -- it is a
Goldstone boson. From this point of view it has  nothing in common with
$\rho$-meson and there are no reasons to expect, that quark distributions in
$\pi$ and $\rho$ are equal. This problem will be resolved and it will be
shown, that valence quark distributions in pion,  longitudinally and
transversally polarized
$\rho$-mesons are quite different.

For reader convenience we present first shortly the method of \cite{11}
and the results for quark distributions in pion. Then valence quark
distributions in longitudinal and transverse $\rho$ mesons are calculated.
The valence quark distributions are reliably calculated in the domain of
intermediate $x$, ~ $0.2 \la x \la 0.65$. The use of the Regge behaviour at
small $x$  and quark counting rules at large $x$ allows one to find the
first and the second moments of quark distributions.

\section{The method}

Consider the non-forward 4-point correlator:

$$
\Pi (p_1, p_2; q, q^{\prime}) = -i \int~ d^4x d^4y d^4z e^{ip_1x + iqy -
ip_2z}
$$
\be
\times \langle 0 \vert T \left \{j^h(x),~ j^{el}(y),~ j^{el}(0),~ j^h(z)
\right \} \vert 0 \rangle
\label{4}
\ee
Here $p_1$ and $p_2$ are the initial and final momenta carried by hadronic
current $j^h$, $q$ and $q^{\prime} = q + p_1 - p_2$ are the initial and
final momenta carried by virtual photons. (Lorenz indeces are omitted). It
will be very essential for us to consider non-equal $p_1, p_2$ and treat
$p^2_1, p^2_2$ as two independent variables. However, we may put $q^2 =
q^{\prime 2} = q^2$ and $t = (p_1 - p_2)^2 = 0$. We are interested in
imaginary part of $\Pi(p^2_1, p^2_2, q^2, s)$ in $s$ channel:

\be
Im \Pi (p^2_1, p^2_2, q^2, s) = \frac{1}{2i} \Biggl [\Pi(p^2_1, p^2_2, q^2,
s + i \varepsilon) - \Pi(p^2_1, p^2_2, q^2, s - i\varepsilon) \Biggr ]
\label{5}
\ee
In order to construct representation of $Im \Pi(p^2_1, p^2_2, q^2, s)$ in
terms of contributions of physical states, let us write for $Im \Pi(p^2_1, p^2_2, q^2, s)$
the double dispersion relation in $p^2_1, p^2_2$:

$$
Im \Pi(p^2_1, p^2_2, q^2, s) = a(q^2, s) + \int \limits^{\infty}_0
~\frac{\varphi(q^2, s, u)}{u - p^2_1} du + \int \limits^{\infty} _{0}
\frac {\varphi (q^2, s, u)}{u - p^2_2} du$$
\be
+ \int \limits ^{\infty}_{0} du_1~ \int \limits ^{\infty} _{0} du_2~ \frac
{\rho(q^2, s, u_1, u_2)}{(u_1 - p^2_1) (u_2 - p^2_2)}
\label{6}
\ee
The second and the third terms in the right-hand side (rhs) of (\ref{6}) may
 be considered as subtraction terms to the last one -- the properly double
 spectral representation. The first term in the rhs of (\ref{6}) is the
 subtraction term to the second and third ones. Therefore, (\ref{6}) has the
 general form of the double spectral representation with one subtraction in
 both variables -- $p^2_1$ and $p^2_2$.  Apply the double Borel
 transformation in $p^2_1, p^2_2$ to (\ref{6}). This transformation kills
 three first terms in rhs of (\ref{6}) and we have

\be
{\cal{B}}_{M^2_1} {\cal{B}}_{M^2_2}~ Im \Pi(p^2_1, p^2_2, q^2, s) = \int
\limits ^{\infty} _{0} du_1~ \int \limits ^{\infty}_{0} du_2 \rho (q^2, s,
u_1, u_2) exp \Biggl [-\frac{u_1}{M^2_1} - \frac{u_2}{M^2_2} \Biggr ]
\label{7}
\ee
The integration region over $u_1, u_2$ may be divided into 4 areas (Fig.1):

I. $u_1 < s_0;~~ u_2 < s_0$

II.$u_1 < s_0; ~~ u_2 > s_0$

III. $u_1 > s_0; ~~ u_2 < s_0$

IV. $u_1, u_2 > s_0$

Using the standard QCD sum rule model of hadronic spectrum and the
hypothesis of quark-hadron duality, i.e. the model with one lowest resonance
plus continuum, one may clearly see, that area I corresponds to resonance
contribution. Spectral density in this area can be written as

\be
\rho(u_1, u_2, x, Q^2) = g^2_h \cdot 2\pi F_2 (x, Q^2) \delta(u_1 - m^2_h)
\delta(u_2 - m^2_h),
\label{8}
\ee
where $g_h$ is defined as

\be
\langle 0 \vert j_h \vert h \rangle = g_h
\label{9}
\ee
(For simplicity we consider the case of the Lorenz scalar hadronic current.
The necessary modifications  for cases of $\pi$ and $\rho$-mesons will be
presented below). If in $Im \Pi(p_1, p_2, q, q^{\prime})$ the structure,
proportional to $P_{\mu} P_{\nu}$ $[P_{\mu} = (p_1 + p_2)_{\mu}/2]$ is
considered, then in the lowest twist approximation $F_2(x, Q^2)$ is the
structure function depending  on the Bjorken scaling variable $x$ and weakly
on $Q^2 = -q^2$.

In area (IV), where both variables $u_{1,2}$ are far from resonance region,
the non-perturbative effects may be neglected, and as usual in sum rules,
the spectral function of hadron state is described by the bare loop spectral
function $\rho^0$ in the same region

\be
\rho(u_1, u_2, x) = \rho^0(u_1, u_2, x)
\label{10}
\ee
In areas (II),(III) one of the variables is far from the resonance region,
but other is in the resonance region, and the spectral function in this
region is some unknown function $\rho = \psi(u_1, u_2, x)$, which
corresponds to transitions like $h \to continuum$ as shown in Fig.2. After
double Borel transformation the physical side of the sum rule can be written
as ($M^2_1, M^2_2$ are Borel mass square)

$$
\hat{B}_1 \hat{B}_2 [Im \Pi] = 2 \pi F_2 (x, Q^2) \cdot g^2_h e^{-m^2_h
(\frac{1}{M^2_1}+ \frac{1}{M^2_2})} + \int \limits^{s_0} _{0} du_1 ~\int
\limits ^{\infty}_{s_0}~ du_2 \psi (u_1, u_2, x) e^{-(\frac{u_1}{M^2_1} +
\frac{u_2}{M^2_2})}
$$
\be
+\int \limits^{\infty} _{s_0} du_1 ~\int
\limits ^{s_0}_0~ du_2 \psi (u_1, u_2, x) e^{-(\frac{u_1}{M^2_1} +
\frac{u_2}{M^2_2})} + \int \limits ^{\infty}_{s_0} \int \limits ^{\infty}
_{s_0} du_1 du_2 \rho^0 (u_1, u_2, x)e^{-(\frac{u_1}{M^2_1} +
\frac{u_2}{M^2_2})}
\label{11}
\ee
In what follows we put $M^2_1 = M^2_2 \equiv 2 M^2$.
(As was shown in \cite{12}, the values of Borel parameters $M^2_1,
M^2_2$ in the double Borel transformation are about twice of
that in the ordinary ones).

One of advantages of this method is that after
double Borel transformation unknown contribution of (II), (III) areas
[the second and third term in (\ref{11})] are exponentially suppressed. Using
duality arguments, we estimate the contribution of all non-resonancee
region (i.e. areas II, III, IV) as a contribution of bare loop in the
same region and demand their value to be small (less than 30\%). So,
equating physical and QCD representation of $\Pi$ and taking into account
cancellation of appropriate parts in the left and right sides, one can
write the following sum rules:

$$
Im~ \Pi^0_{QCD} + \mbox{Power~ correction} = 2 \pi F_2 (x, Q^2) g^2_h e
^{-m^2_h(\frac{1}{M^2_1} + \frac{1}{M^2_2})}
$$
\be
Im \Pi^0_{QCD} = \int \limits^{s_0}_{0}\int \limits^{s_0}_{0}\rho^0 (u_1,
u_2, x) e^{- \frac{u_1 + u_2}{2M^2}}
\label{12}
\ee
It can be shown (see below), that for box
diagram $\psi(u_1, u_2, x)) \sim \delta(u_1 - u_2)$, and, as a consequence,
the second and third terms in (\ref{11}) are zero in our model of hadronic
spectrum.

It is worth mentioning that if we would consider the forward
scattering amplitude from the beginning, put $p_1 = p_2 = p$ and perform
Borel transformation in $p^2$, then unlike (\ref{11}), the contributions of
the second and third terms in (\ref{6}) would not be suppressed comparing
with the interesting for us lowest resonance contribution. They just
correspond to the non-diagonal transition matrix elements discussed in the
Introduction and are proportional to

\be
\langle 0 \vert j^h \vert h^{\ast} \rangle ~ \frac{1}{p^2 - m^{*2}_h}~
\langle h^{\ast} \vert j^{el}(x) j^{el} (0) \vert h \rangle ~ \frac{1}{p^2 -
m^2_h} \langle h \vert j^h \vert 0 \rangle
\label{13}
\ee
From decomposition

\be
\frac{1}{p^2 - m^{*2}_h} ~ \frac{1}{p^2 - m^2_h} = \frac{1}{m^{*2}_h -
m^2_h} \Biggl (\frac{1}{p^2 - m^{*2}_h} - \frac{1}{p^2 - m^2_h} \Biggr
)
\label{14}
\ee
it is clear that in this case (\ref{13}) may contribute to the second (or
third) term in (\ref{6}) and after Borel transformation the contribution of
the second term in (\ref{14}) has the same Borel exponent $e^{-m^2_h/M^2}$
as the lowest resonance contribution. The only difference is in pre-exponent
factors: they are $1/M^2$ in front of the resonance term and Const. in front
of non-diagonal terms. This difference was used in order to get rid of
non-diagonal terms:  application of the differential operator
$(\partial/\partial (1/M^2)e^{m^2_h/M^2}$ to the sum rule kills the Borel
non-suppressed nondiagonal terms, but deteriorates the accuracy and shrinks
the applicability domain of the sum rule (particularly, the domain in $x$,
where the sum rule is valid).

Show now that in the used here model of hadronic spectrum--resonance plus
continuum, where continuum is given by the bare loop contribution, the
second and third terms in (\ref{11}) are in fact zero. Consider the bare loop
represented by the diagram of Fig.3. For simplicity restrict ourselves by the
case when all propagators are bosonic and all currents are scalar. (In the
realistic case with quarks in internal lines conclusion will be the
same). The imaginary part in $s$-channel of Fig.3 diagram is given by (quark
masses are neglected):

\be
Im~ T(p^2_1, p^2_2, q^2, s) = \int ~ d^4k \frac{1}{(p_1 - k)^2 (p_2 -
k)^2} \delta \Biggl [(p_1 + q_1 - k)^2 \Biggr ] \delta (k^2)
\label{15}
\ee
Neglecting the higher twist terms $\sim p^2_{1,2}/q^2$ (\ref{15}) is equal to
(see \cite{11}, Appendix)

\be
Im~ T(p^2_1, p^2_2, q^2, s) =  \frac{\pi}{4 \nu x}~ \int \limits ^{\infty}
_{0}  \frac{1}{(u - p^2_1)(u - p^2_2)} ~du
\label{16}
\ee
where $\nu = qP$ and $x = -q^2/2 \nu$. To derive (\ref{16}) it is convenient
to introduce

$$
P = (p_1 + p_2)/2
$$
\be
r = p_1 - p_2, ~ r^2 = 0
\label{17}
\ee
and in the Lorenz system, where 4-vector $P$ has only $z$-component and
$q$ - only time and $z$-components

\be
r_0 \approx r_z \approx \frac{1}{2}~ \frac{p^2_1 -  p^2_2}{\sqrt{- P^2}}, ~~
r^2_{\bot} = -\frac{1}{4}~ \frac{q^2(p^2_1 - p^2_2)^2}{\nu^2 - q^2 P^2},
\label{18}
\ee
$r^2_{\bot}$  is of higher twist and may be neglected. As follows from
(\ref{16}), in case of bare loop of Fig.3 the spectral function is
proportional to $\delta(u_1 - u_2)$ and the contributions of areas II, III
in Fig.1 are zero. Since one may expect, that more complicated diagrams are
smaller and in any case their contributions in domains II, III are
suppressed by Borel exponents, one may safely neglect them, as was done in
(\ref{12}).


\section{ Quark distributions in pion}

It is enough to find the distribution of valence $u$-quark in $\pi^+$, since
$\bar{d}(x) = u(x)$. The most suitable hadronic current in this case is the
axial current

\be
j_{\mu 5} = \bar{u} \gamma_{\mu} \gamma_5 d
\label{19}
\ee
In order to find the $u$-quark distribution, the electromagnetic current is
chosen as $u$-quark current with the unit charge

\be
j^{el}_{\mu} = \bar{u} \gamma_{\mu} u
\label{20}
\ee
The bare loop Fig.3 contribution is given by

$$
Im~ \Pi_{\mu \nu \lambda \sigma} = -\frac{3}{(2 \pi)^2}~ \frac{1}{2}~ \int~
\frac{d^4 k}{k^2} ~ \frac{1}{(k + p_2 - p_1)^2} \delta [(q + k)^2
 ] \delta [p_1 - k)^2 ]
$$
\be
\times Tr \left \{\gamma_{\lambda} \hat{k} \gamma_{\mu}(\hat{k} + \hat{q})
\gamma_{\nu} (\hat{k} + \hat{p}_2 - \hat{p}_1) \gamma_{\sigma} (\hat{k} -
\hat{p}_1) \right \}
\label{21}
\ee
The tensor structure, chosen to construct the sum rule is a structure
proportional to $P_{\mu} P_{\nu} P_{\lambda} P_{\sigma}/\nu$.  The reasons
are the following. As is known, the results of the QCD sum rules
calculations are more reliable, if invariant amplitude at kinematical
structure with maximal dimension is used. Different $p_1 \not=p_2$ are
important for us only in denominators, where they allow one to separate the
terms in dispersion relations. In numerators one may restrict oneself to
consideration of terms proportional to 4-vector $P_{\mu}$ and ignore the
terms $\sim r_{\mu}$. Then the factor $P_{\mu} P_{\nu}$ provides the
contribution of $F_2(x)$ structure function and the factor $P_{\lambda}
P_{\sigma}$ corresponds to contribution of spin zero states. (The factor
$1/\nu$ is scaling \\
factor : $w_2 = F_2/\nu$.)

Let us use the notation

\be
\Pi_{\mu \nu \lambda \sigma} = (P_{\mu} P_{\nu} P_{\lambda} P_{\sigma}/\nu)
\Pi (p^2_1, p^2_2, x) + ...
\label{22}
\ee
Then $Im \Pi (p^2_1, p^2_2, x)$ can be calculated from (\ref{21}) (eq.'s
(\ref{15}),(\ref{16}) are exploited) and the result is \cite{11}:

\be
Im \Pi (p^2_1, p^2_2, x) = \frac{3}{\pi} x^2 (1 - x) \int \limits ^{\infty}
_{0}~ du_1 \int \limits ^{\infty} _{0} du_2 ~ \frac{\delta(u_1 - u_2)}{(u_1
- p^2_1)(u_2 - p^2_2)}
\label{23}
\ee
The matrix element of the axial current between vacuum and pion state is
well known

\be
\langle 0 \vert j_{\mu 5} \vert \pi \rangle = i p_{\mu} f_{\pi}
\label{24}
\ee
where $f_{\pi}= 131 MeV$. The use of (\ref{12}), (\ref{23}), (\ref{24})
gives the sum rule for valence $u$-quark distribution in pion in the bare
loop approximation \cite{11}:

\be
u_{\pi}(x) = \frac{3}{2 \pi^2}~\frac{M^2}{f^2_{\pi}} x (1-x) (1 -
e^{-s_0/M^2}) e^{m^2_{\pi}/M^2},
\label{25}
\ee
where $s_0$ is the continuum threshold.
In ref.\cite{11} the following corrections to (\ref{25}) were accounted:

1. Leading order (LO) perturbative corrections, proportional to
$ln(Q^2/\mu^2)$ , where $\mu^2$ is the normalization point. In what follows
the normalization point will be chosen to be equal to the Borel parameter
$\mu^2 = M^2$.

2.  Power corrections - higher order terms of OPE. Among the latter, the
dimension-4 correction, proportional to gluon condensate $\langle 0 \vert
\frac{\alpha_s}{\pi} G^n_{\mu \nu}~ G^n_{\mu\nu} \vert 0 \rangle$ was first
accounted, but it was found that the gluon condensate contribution to the
sum rule vanishes after double borelization. There are two types of vacuum
expectation values (v.e.v) of dimension 6: one, where only gluonic fields
enter:

\be
\frac{g_s}{\pi} \alpha_s f^{abc} \langle 0 \vert G^{a}_{\mu \nu}~ G^b_{\nu
\lambda}~ G^c_{\lambda \mu} \vert 0 \rangle
\label{26}
\ee
and the other, proportional to four-quark operators

\be
\langle 0 \vert \bar{\psi} \Gamma \psi \cdot \bar{\psi} \Gamma \psi \vert 0
\rangle
\label{27}
\ee
It was shown in \cite{11} that terms of the first type cancel in the sum
rule and only terms of the second type survive. For the latter one may use
the factorization hypothesis which reduces all the terms of this type to the
square of quark condensate.

 A remark is in order here. As was mentioneed in the Introduction, the
 present approach is invalid at small and large $x$. No-loop 4-quark
 condensate contributions, like Fig.4, are proportional to $\delta(1-x)$ and
 being outside of the applicability domain of the approach, cannot be
 accounted. In the same way, the diagrams, which can be considered as a
 radiative corrections to those, proportional to $\delta(1-x)$, must be also
 omitted.

 All dimension-6 power corrections to the sum rule were calculated in
 ref.\ref{12} and the final result is given by (the pion mass is neglected):

$$
xu_{\pi}(x) = \frac{3}{2\pi^2}\frac{M^2}{f^2_{\pi}}x^2(1-x)
\Biggl [ \Biggl ( 1+ \Biggl (\frac{a_s(M^2)\cdot ln(Q^2_0/M^2)}{3\pi}\Biggr
)$$

$$\times \Biggl ( \frac{1+4x ln(1-x)}{x}- \frac{2(1-2x)ln x}{1-x}\Biggl )
\Biggl )\cdot (1-e^{-s_0/M^2}) $$

\be
-\frac{4\pi \alpha_s(M^2)\cdot 4\pi \alpha_s a^2}{(2\pi)^4 \cdot
3^7\cdot 2^6\cdot M^6} \cdot \frac{\omega(x)}{x^3(1-x)^3}\Biggr ],
\label{28}
\ee
where $\omega(x)$ is 4-order polynomial in $x$,

\be a =
-(2 \pi)^2 \langle 0 \vert  \bar{\psi} \psi \vert 0 \rangle
\label{29}
\ee

$$
\omega(x) = -5784x^4 - 1140x^3 - 20196x^2
$$
$$
+ 20628x - 8292) ln(2) + 4740x^4 + 8847x^3
$$
\be
+ 2066x^2 - 2553x + 1416
\label{30}
\ee
$u_{\pi}(x)$ may be used as an initial condition at $Q^2 = Q^2_0$
for solution of
QCD evolution equations (DGLAP) equations \footnote{There was a misprint in
the corresponding equation in \cite{11} (eq.\ref{40} of \cite{11}): instead
of $\alpha_s(M^2)$ in the last term was $\alpha_s(Q^2_0)$. In numerical
calculations the correct value $\alpha_s(M^2)$ was taken.}.

In numerical calculations we choose: the effective LO QCD parameter
$\Lambda^{LO}_{QCD} = 200 MeV,~ Q^2_0 = 2 GeV^2, ~ \alpha_s a^2 (1 GeV^2) =
0.13 GeV^6$ \cite{11}. The continuum threshold was varied in the interval
$0.8 < s_0 < 1.2 GeV^2$ and it was found, that the results only slightly
depend on it. The analysis of the sum rule (\ref{28}) shows, that it is
fulfilled in the region $0.15 < x < 0.7$; the power corrections are less
than 30\% and the continuum contribution is small ($< 25$\%). The stability
in the Borel mass parameter $M^2$ dependence in the region $0.4 GeV^2 < M^2
< 0.6 GeV^2$ is good. The result of our calculation of valence distribution
in pion $x u_{\pi}(x,Q^2_0)$ is shown  in Fig.5.

Suppose, that at small $x \la 0.15 ~~ u_{\pi}(x) \sim 1/\sqrt{x}$ according
to Regge behaviour and at large $x \ga 0.7~~ u_{\pi}(x) \sim (1-x)^2$
according to quark counting rules. Then, matching these functions with
(\ref{28}), one may find the numerical values of the first and second
moments of $u$-quark distribution

\be
{\cal{M}}_1 = \int \limits^1 _0 u_{\pi} (x) dx \approx 0.84 ~~ (0.85)
\label{31}
\ee

\be
{\cal{M}}_2 = \int \limits^1_0 xu_{\pi} (x) dx \approx 0.21 ~~ (0.23)
\label{32}
\ee
(In the parenthesis we give the values, corresponding to $u_{\pi}(x) \sim 1
- x$ behaviour at large $x$.) The results only slightly depend on the points
of matching (not more than 5\%, when the lower matching point is varied in
the region 0.15 - 0.2 and the upper one in the region 0.65 - 0.75).
${\cal{M}}_1$ has the meaning of the number of $u$-quarks in $\pi^+$ and
should be ${\cal{M}} = 1$. The deviation of (\ref{31}) from 1 characterizes
the accuracy of our calculation. ${\cal{M}}_2$ has the meaning of the part
of pion momentum carried by valence $u$-quark. Therefore, valence $u$ and
$\bar{d}$ quarks are carrying about 40\% of the total momentum. In Fig.5 we
plot also the valence $u$-quark distribution found in \cite{6} by
fitting the data on production of $\mu^+\mu^-$ and $e^+e^-$ pairs in
pion-nucleon collisions (Drell-Yan process). Comparing with the found here
distribution it must be taken in mind, that the accuracy of our curve is of
order of $ 10 - 20\%$, the last number refers to the border of the
applicability domain. $U$-quark distribution found from the data on the
Drell-Yan process is also not free from uncertainties. Particularly, what is
measured in the Drell-Yan process is the quark fragmentation function into
pion defined at $q^2 > 0$. In order to get quark distribution in pion
defined at $q^2 < 0$, the procedure of analytical continuation is used,
which may introduce some uncertainties, especially, at low normalization
point, like $Q^2_0 = 2~ GeV^2$, to which the data in Fig.5
 refer. For all these reasons we consider the agreement of
two curves as good. The calculation of valence $u$-quark distribution in
pion in the instanton model was done recently \cite{13}. At intermediate $x$
the values of $xu_{\pi}(x)$ found in \cite{13}, are about 20\%
higher, in comparison with ours.
Recently also the pions valence quark momentum distribution using a
Dyson-Schwinger equation model was found in \cite{cd}.
Our results are in reasonable agreement
with the results of  \cite{cd}.

\section{Quark distributions in $\rho$-meson}

Let us calculate valence $u$-quarks distribution in $\rho^+$-meson ($\rho$
-- width is neglected). The choice of hadronic current is evident

\be
j_{\mu}^{\rho} = \overline{u}\gamma_{\mu}d
\label{33}
\ee
The matrix element $\langle \rho^+\mid j^{\rho}_{\mu}\mid 0 \rangle$ is
given by

\be
\langle \rho^+\mid j^{\rho}_{\mu}\mid 0 \rangle =
\frac{m^2_{\rho}}{g_{\rho}}e_{\mu}
\ee
where $m_{\rho}$ is the $\rho$-meson mass, $g_{\rho}$ is the $\rho-\gamma$
coupling constant, $g^2_{\rho}/4\pi=1.27,~e_{\mu}$  is the $\rho$-meson
polarization vector. Consider the coordinate system, where the collision of
$\rho$-meson with momentum $p$ and virtual photon with momentum $q$
proceeds along $z$-axes. The averaging over $\rho$ polarizations is given by
the formulae:\\
1. Longitudinally polarized $\rho$:

\be
e^L_{\mu}e^L_{\nu}=
\Biggl ( q_{\mu} - \frac{\nu p_{\mu}}{m^2_{\rho}}\Biggr )
\Biggl ( q_{\nu} - \frac{\nu p_{\nu}}{m^2_{\rho}}\Biggr )
\frac{m^2_{\rho}}{\nu^2-q^2m^2_{\rho}}
\label{35}
\ee
2. Transversally polarized $\rho$:

\be
\sum_{T,r}e^r_{\mu}e^r_{\nu} = -\Biggl ( \delta_{\mu\nu} -
\frac{p_{\mu}p_{\nu}}{m^2_{\rho}} \Biggr ) -
\frac{m^2_{\rho}}{\nu^2-q^2m^2_{\rho}}
\Biggl ( q_{\mu}- \frac{\nu p_{\mu}}{m^2_{\rho}} \Biggr )
\Biggl ( q_{\nu}- \frac{\nu p_{\nu}}{m^2_{\rho}} \Biggr )
\label{36}
\ee
The imaginary part of the forward $\rho-\gamma$  scattering amplitude
$W_{\mu\nu\lambda\sigma}$
(before multiplication by $\rho$-polarizations)  satisfies the equations:
$W_{\mu\nu\lambda\sigma} q_{\mu}=$ $W_{\mu\nu\lambda\sigma} q_{\nu}=$
$W_{\mu\nu\lambda\sigma} p_{\lambda}=$ $W_{\mu\nu\lambda\sigma}
p_{\sigma}=0$, which follows from current conservation. (The indeces
$\mu,\nu$ refer to initial and final photon; $\lambda,\sigma$ -- to initial
and final $\rho$.)  The general form of $W_{\mu\nu\lambda\sigma} $  is:

$$W_{\mu\nu\lambda\sigma} = \Biggl [\Biggl
( \delta_{\mu\nu} - \frac{q_{\mu}q_{\nu}}{q^2}\Biggr ) \Biggl (
\delta_{\lambda\sigma}-\frac{p_{\lambda}p_{\sigma}}{m^2_{\rho}}\Biggr ) A-
\Biggl ( \delta_{\mu\nu} - \frac{q_{\mu}q_{\nu}}{q^2}\Biggr ) \Biggl (
q_{\lambda}- \frac{\nu p_{\lambda}}{m^2_{\rho}}\Biggr ) \Biggl (
q_{\sigma}-\frac{\nu p_{\sigma}}{m^2_{\rho}}\Biggr ) B$$

$$ - \Biggl ( p_{\mu} - \frac{\nu q_{\mu}}{q^2}\Biggr )
\Biggl ( p_{\nu} - \frac{\nu q_{\nu}}{q^2}\Biggr ) \Biggl (
\delta_{\lambda\sigma} -\frac{p_{\lambda} p_{\sigma}}{m^2_{\rho}} \Biggr ) C
+ \Biggl ( p_{\mu} - \frac{\nu q_{\mu}}{q^2}\Biggr )
\Biggl ( p_{\nu} - \frac{\nu q_{\nu}}{q^2}\Biggr )$$

\be
\times \Biggl ( q_{\lambda} - \frac{\nu p_{\lambda}}{m^2_{\rho}}\Biggr )
\Biggl ( q_{\sigma} - \frac{\nu p_{\sigma}}{m^2_{\rho}}\Biggr ) D \Biggr ]
\label{37}
\ee
where A,B,C,D are  invariant functions. Average (\ref{37}) over
polarizations for longitudinal and transverse $\rho$-mesons. We have

$$W_{\mu\nu\lambda\sigma} e^L_{\lambda}e^L_{\sigma} = -\Biggl (
\delta_{\mu\nu} - \frac{q_{\mu}q_{\nu}}{q^2}\Biggr ) \Biggl ( A +
\frac{\nu^2 -q^2m^2_{\rho}}{m^2_{\rho}}B \Biggr ) + $$

\be
+ \Biggl ( p_{\mu} - \nu\frac{q_{\mu}}{q^2}\Biggr )
\Biggl ( p_{\nu} - \nu\frac{q_{\nu}}{q^2}\Biggr ) \Biggl ( C +
\frac{\nu^2-q^2m^2_{\rho}}{m^2_{\rho}} D \Biggr )
\label{38}
\ee

\be
\frac{1}{2}\sum_{T,r} W_{\mu\nu\lambda\sigma} e^r_{\lambda}e^r_{\sigma}
=-\Biggl ( \delta_{\mu\nu} - \frac{q_{\mu}q_{\nu}}{q^2} \Biggr ) A + \Biggl
(p_{\mu} -\frac{\nu q_{\mu}}{q^2}\Biggr )
\Biggl (p_{\nu} -\frac{\nu q_{\nu}}{q^2}\Biggr ) C
\label{39}
\ee
From comparison of (\ref{37}) and (\ref{38}), (\ref{39}) it is clear, that
the proportional to $p_{\mu}p_{\nu}$ structure function $F_2(x)$ in the
scaling limit $(\nu^2 \gg \mid q^2 \mid m^2_{\rho})$ is given by the
contribution of invariants $C+(\nu^2/m^2_{\rho})D$  in case of longitudinal
and by invariant $C$  in case of transversal $\rho$-mesons. This means, that
in the forward scattering amplitude $W_{\mu\nu\lambda\sigma}$ (\ref{37}) one
must separate the structure proportional to
$p_{\mu}p_{\nu}p_{\lambda}p_{\sigma}$ in the first case and the structure
$\sim p_{\mu}p_{\nu}\delta_{\lambda\sigma}$ in the second case.

Consider now the non-forward 4-point correlator

$$\Pi_{\mu\nu\lambda\sigma} (p_1,p_2;q,q^{\prime}) = -i\int d^4 x d^4 y d^4
ze^{ip_1x+iqy-ip_2z} $$

\be
\times \langle 0\mid T~\{ j^{\rho}_{\lambda}
(x),~j^{el}_{\mu}(y),~j^{el}_{\nu}(0),~j^{\rho}_{\sigma}(z)\}\mid 0
\rangle,
\label{40}
\ee
where the currents $j^{el}_{\mu}(x)$ and $j^{\rho}_{\lambda}(x)$ are given
by (\ref{20}) and (\ref{33}). It is evident from the said above, that in the
non-forward amplitude for determination of $u$-quark distribution in
longitudinal $\rho$-meson the most suitable tensor structure is that,
proportional to $P_{\mu} P_{\nu} P_{\sigma}P_{\lambda}$, while $u$-quark
distribution in transverse $\rho$  can be found by considering the invariant
function at the structure $-P_{\mu}P_{\nu}\delta_{\lambda\sigma}$ ($P$ is
given by (\ref{17})).

Calculate first the bare loop  contribution (diagram of Fig.3). In case of
longitudinal $\rho$-meson the tensor structure, which is separated  is the
same, as in the case of pion. Since at $m_q=0$  bare loop contributions for
vector and axial hadronic currents are equal, the only difference from the
pion case is in the normalization. It can be shown, that $u$-quark
distribution in longitudinal $\rho$-meson can be found from (\ref{25}) by
substitutions $m_{\pi}\to m_{\rho}$, $f_{\pi}\to m_{\rho}/g_{\rho}$ and
therefore

\be
u^L_{\rho} = \frac{3}{2\pi^2}\frac{M^2}{m^2_{\rho}} g^2_{\rho} x
(1-x)e^{m^2_{\rho}/M^2}(1-e^{-s_0/M^2})
\label{41}
\ee
The calculation of $g^2_{\rho}$  performed in the same approximation in
\cite{14}  leads to

\be
\frac{g^2_{\rho}M^2}{m^2_{\rho} 4\pi}(1-e^{-s_0/M^2}) e^{m^2_{\rho}/M^2} =
\pi \label{42}
\ee
(the definition of $g_{\rho}$ used here differs from that
in \cite{14} by a factor $1/\sqrt{2}$). The substitution of (\ref{42}) in
(\ref{41}) gives

\be
u^L_{\rho} (x) = 6x(1-x)
\label{43}
\ee
and we have

\be
\int\limits^1_0 u^L_{\rho}(x) dx = 1,
\label{44}
\ee
as it should be. Also,

\be
\int\limits^1_0 xu^L_{\rho} (x) dx =\frac{1}{2},
\label{45}
\ee
is in correspondence with naive quark model, where $u$-quarks are carrying
one half of total momentum. The calculation of the term, proportional to the
structure $P_{\mu}P_{\nu}\delta_{\lambda\sigma}$ in the diagram of Fig.3
gives

\be
Im \Pi^{(0)}_{\mu\nu\lambda\sigma}\equiv -\frac{1}{\nu}P_{\mu}P_{\nu}
\delta_{\lambda\sigma} Im \Pi^{(0)}_T = -\frac{3}{2\pi}\frac{1}{\nu}
P_{\mu}P_{\nu}\delta_{\lambda\sigma}x \Biggl [\frac{1}{2}- x (1-x)\Biggr ]
\int \frac{udu}{(u-p^2_1)(u-p^2_2)}
\label{46}
\ee
After borelization we get for $u$-quark distribution in transversally
polarized $\rho$-meson in bare loop approximation

\be
u^T_{\rho}(x) = \frac{3}{(2\pi)^2}
g^2_{\rho}\frac{M^4}{m^4_{\rho}}e^{m^2_{\rho}/M^2} E_1\Biggl
(\frac{s_0}{M^2}\Biggr ) \Biggl [ \frac{1}{2} - x (1-x)\Biggl ]
\label{47}
\ee
where

\be
E_1(z) = 1 - (1+z)e^{-z}
\label{48}
\ee
Let us use (\ref{42}), put $M^2=m^2_{\rho}$ and neglect the terms $\sim
e^{-s_0/m^2_{\rho}}$. Then a simple formula for $u^T_{\rho}(x)$ follows:

\be
u^T_{\rho}(x) = 3 \Biggl [ \frac{1}{2} - x (1-x)\Biggr ]
\label{49}
\ee
$U$-quark  distribution (\ref{49}) has the expected properties:

\be
\int\limits^1_0 u^T_{\rho}(x) dx = 1
\label{50}
\ee
 \be
\int\limits^1_0 xu^T_{\rho}(x) dx = \frac{1}{2}
\label{51}
\ee

Take in account LO  perturbative correction, proportinal to $ln~Q^2/\mu^2$
and choose $Q^2=Q^2_0$ for the point where we
calculate our sum rules. The result is
(the second term in square brackets  corresponds to the perturbative
correction):

$$u_{\rho}^L(x) = \frac{3M^2}{4\pi^2}
\frac{g^2_{\rho}}{m^2_{\rho}}e^{m^2_{\rho}/M^2} x(1-x)\Biggl [ 1 +
\frac{a_s(\mu^2)ln(Q^2_0/\mu^2)}{3\pi}$$

\be
\times \Biggl ( 1/x + 4ln (1-x) - \frac{2(1-2x)ln~x}{1-x}\Biggr ) \Biggr
] (1-e^{-s_0/M^2})
\label{52}
\ee
and

$$u^T (x) = \frac{3}{8\pi^2}
\frac{g^2_{\rho}}{m^4_{\rho}}e^{m^2_{\rho}/M^2}\cdot M^4 \cdot E_1\Biggl
(\frac{s_0}{M^2}\Biggr ) \cdot \varphi_0(x)$$

\be
\Biggl [ 1 + \frac{ln(Q^2_0/\mu^2)\cdot \alpha_s(\mu^2)}{3\pi}\cdot
\Biggl (
(4x-1)/\varphi_0(x) + 4ln(1-x) - \frac{2(1-2x+4x^2)lnx}{\varphi_0(x)}\Biggr )
\Biggr ]
\label{53}
\ee
where

\be
\varphi_0(x)=1-2x(1-x)
\label{54}
\ee

Turn now to consideration of power correction
contribution to the sum rules. The power correction of lower dimension is
proportional to the gluon condensate
$\langle G^q_{\mu\nu}G^q_{\mu\nu}\rangle$ with $d=4$. As was discussed above, only $s$-shannel
diagrams (Fig.1)  exist in the case of double borelization. The
$\langle G^q_{\mu\nu}G^q_{\mu\nu}\rangle$ correction was calculated in a
standard way  in the Fock-Schwinger gauge
$x_{\mu}A_{\mu}=0$ \cite{15}.

The quark propagator $iS(x,y)=\langle \psi(x)\psi(y)\rangle$ in the external
field $A_{\mu}$ has the well-known form \cite{12,8,16}. In the
$\pi$-meson case the sum of all diagrams, corresponding to $\langle
G^a_{\mu\nu}G^a_{\mu\nu}\rangle$  corrections (Fig.6), was found to be zero
after double borelization \cite{11}.
For $\rho$-meson, however, $\langle G^a_{\mu\nu}G^a_{\mu\nu}\rangle$
correction are equal to zero for longitudinally polarized $\rho$ ($\rho_L$)
but are not vanishing for transversally polarized $\rho(\rho_T)$

\be
Im\Pi^{(d=4)}_T = -\frac{\pi}{8x}\langle 0\mid\frac{\alpha_s}{\pi}
G^2_{\mu\nu}\mid 0\rangle
\label{55}
\ee
All diagrams here and in what follows are calculated using a program for
analytical calculations REDUCE. Before we discuss $d=6$  contribution, let
us remind, that we should not take into account non-loop diagrams and
diagrams, which can be treated  as their evolution (for example see Fig.7).
There is a large number of loop diagrams for $d=6$ correction. It is
convenient to divide  them into two types and discuss these types
separately: \\
Type I-diagrams where only interaction with the external gluon
field is taken into account -- see Fig.8,9.\\
Type II -- diagrams, where expansion of quark field is also accounted
($\nabla$-covariant derivate)

\be
\psi(x) = \psi(0) + x_{\alpha_1}~[\nabla_{\alpha_1}\psi(0)] +
\frac{1}{2}x_{\alpha_1}x_{\alpha_2}~[\nabla_{\alpha_1} \nabla_{\alpha_2}
\psi(0)]+ ...
\label{56}
\ee
The examples of such diagrams are shown in Fig.10,11. Discuss briefly the
special features of calculation of this two types of diagrams. The diagram
of type I (Fig.8) are, obviously, proportional  to $\langle 0 \mid g^3
f^{abc}G^a_{\mu\nu}G^b_{\alpha\beta}G^c_{\rho\sigma} \mid 0 \rangle $ and
when calculating it is convenient to use the representation of this tensor
structure suggested in \cite{17}

\be
\langle 0 \mid g^3 f^{abc}G^a_{\mu\nu}G^b_{\alpha\beta}G^c_{\rho\sigma}
\mid 0 \rangle
=1/24
\langle 0 \mid g^3
f^{abc}G^a_{\gamma\delta}G^b_{\delta\epsilon}G^c_{\epsilon\gamma}
\mid 0 \rangle
\label{57}
\ee

$$\times (g_{\mu\sigma} g_{\alpha\nu} g_{\beta\rho} +
g_{\mu\beta}g_{\alpha\rho}g_{\sigma\nu} + g_{\alpha\sigma} g_{\mu\rho}
g_{\nu\beta} + g_{\rho\nu}g_{\mu\alpha}g_{\beta\sigma}$$

$$ - g_{\mu\beta}g_{\alpha\sigma}g_{\rho\nu} -
g_{\mu\sigma}g_{\nu\beta}g_{\alpha\rho} -
g_{\alpha\nu}g_{\mu\rho}g_{\beta\sigma} -
g_{\beta\rho}g_{\mu\alpha}g_{\nu\sigma}) $$
The diagrams of Fig.9 are proportional to
$\langle 0 \mid
D_{\rho}G^a_{\mu\nu}D_{\tau}G^a_{\alpha\beta}\mid 0\rangle$
and  $\langle 0 \mid
G^a_{\mu\nu}D_{\rho}D_{\tau}G^a_{\alpha\beta}\mid 0\rangle$. Using the
equation of motion it was found in \cite{17} that\footnote{Note that our
choice of sign of $g$ is opposite to those of \cite{17}.}

$$-\langle 0 \mid D_{\rho}G^a_{\mu\nu}D_{\sigma}G^a_{\alpha\beta} \mid 0
\rangle =
\langle 0 \mid G^a_{\mu\nu}D_{\rho}D_{\sigma}G^a_{\alpha\beta} \mid 0
\rangle
=2O^-~[g_{\rho\sigma}(g_{\mu\beta}g_{\alpha\nu}-g_{\mu\alpha}g_{\nu\beta})$$

$$+\frac{1}{2}(g_{\mu\beta}g_{\alpha\sigma}g_{\rho\nu}+g_{\alpha\nu}
g_{\mu\rho}g_{\beta\sigma} -
g_{\alpha\sigma}g_{\mu\rho}g_{\nu\beta}-g_{\rho\nu}g_{\mu\alpha}
g_{\beta\sigma})]$$

$$+ O^+(g_{\mu\sigma}g_{\alpha\nu}g_{\beta\rho} +
g_{\mu\beta}g_{\alpha\rho}g_{\sigma\nu} -
g_{\mu\sigma}g_{\alpha\rho}g_{\nu\beta} -
g_{\rho\beta}g_{\mu\alpha}g_{\nu\sigma}),$$

\be
O^{\pm} = \frac{1}{72}\langle 0 \mid g^2j^a_{\mu}j^a_{\mu}\mid 0 \rangle \mp
\frac{1}{48} \langle 0 \mid g f^{abc} G^a_{\mu\nu} G^b_{\nu\lambda}
G^c_{\lambda\mu} \mid 0\rangle,
\label{58}
\ee
where $j^a_{\mu}=\sum \bar{\psi}_i\gamma_{\mu}(\lambda^a/2)\psi_i$.

From (\ref{57}) and (\ref{58}) one may note that these tensor structures are
proportional to two vacuum averages:

$$\langle 0 \mid g^2j^2_{\mu} \mid 0 \rangle ~~~\mbox{and}~~~
\langle 0 \mid g^3 G^a_{\mu\nu} G^b_{\nu\rho} G^c_{\rho\mu}f^{abc} \mid 0
\rangle.$$
The first of these, $\langle 0 \mid g^2j^2_\mu\mid 0\rangle$, by use of the
factorization hypothesis easily reduces to $\langle g\bar{\psi}\psi\rangle^2$
which is well known,

\be
\langle 0 \mid g^2j^2_{\mu}\mid 0 \rangle = -(4/3)[\langle 0 \mid
g\bar{\psi}\psi \mid 0 \rangle ]^2.
\label{59}
\ee
But $\langle 0 \mid g^3
G^a_{\mu\nu} G^b_{\nu\rho} G^c_{\rho\mu}f^{abc} \mid 0 \rangle$ is not well
known; there are only some estimates based on the instanton model
\cite{18,19}. In the $\pi$-meson case the terms, proportional to\\
$\langle 0\mid  g^3 f^{abc} G^a_{\mu\nu}G^b_{\nu\rho}$ $G^c_{\rho\mu} \mid 0
\rangle$  exactly cancelled \cite{11}. The similar cancellation takes place
for $\rho_L$. But there is no such cancellation for $\rho_T$ and one should
estimate $\langle 0 \mid g^3
f^{abc}G^a_{\mu\nu}G^b_{\nu\rho}G^c_{\rho\mu}\mid 0 \rangle$. The
estimation, based on the instanton model \cite{18}, gives

\be
-\langle g^3 f^{abc}G^a_{\mu\nu} G^b_{\nu\rho} G^c_{\rho\mu}\rangle =
\frac{48\pi^2}{5} \frac{1}{\rho^2_c} \langle 0\mid(\alpha_s/\pi) G^2_{\mu\nu}
\mid 0\rangle,
\label{60}
\ee
where $\rho_c$  is the effective instanton
radius.

Among the diagrams of type II (Figs.10,11) only those, where the
interaction with the vacuum takes place inside the loop, are considered. Such
diagrams cannot be treated as the evolution of any non-loop diagrams and are
pure power corrections of dimension 6. All these diagrams are, obviously,
proportional to

$$\langle 0 \mid \bar{\psi}^d_{\alpha}\psi^b_{\beta}D_{\rho}G^n_{\mu\nu}\mid
0 \rangle,~~~\langle 0 \mid
\bar{\psi}^d_{\alpha}(\nabla_{\tau}\psi^b_{\beta})G^n_{\mu\nu} \mid 0
\rangle,~~~\langle 0 \mid
(\nabla_{\tau}\bar{\psi}^d_{\alpha})\psi^b_{\beta}G^n_{\mu\nu}\mid 0
\rangle.$$
These tensor structure were considered in \cite{12} where with the help of
equation of motion the following results were obtained\footnote{The sign
errors in front of $g$ in eq.'s (\ref{23}),(\ref{24}),(A1),(A2) of \cite{12}
are corrected.}:

\be
\langle 0 \mid \bar{\psi}^d_{\alpha}\psi^b_{\beta}(D_{\sigma}G_{\mu\nu})^n
\mid 0 \rangle = \frac{g\langle 0 \mid \bar{\psi}\psi\mid 0
\rangle^2}{3^3\cdot 2^5} (g_{\sigma\nu}\gamma_{\mu} -
g_{\sigma\mu}\gamma_{\nu})_{\beta\alpha}(\lambda^n)^{bd},
\label{61}
\ee

\be
\langle 0 \mid \bar{\psi}^d_{\alpha}(\nabla_{\sigma}\psi_{\beta})^b
G^n_{\mu\nu}\mid 0 \rangle = \frac{g\langle 0 \mid \bar{\psi}\psi\mid 0
\rangle^2}{3^3\cdot 2^6}
[g_{\sigma\mu}\gamma_{\nu}-g_{\sigma\nu}\gamma_{\mu} - i
\varepsilon_{\sigma\mu\nu\lambda}\gamma_5\gamma_{\lambda}]_{\beta\alpha}
(\lambda^n)^{bd}.
\label{62}
\ee
The term
$\langle 0 \mid (\nabla_{\sigma}\bar{\psi}_{\alpha})^d
\bar{\psi}^b_{\beta}G^n_{\mu\nu}\mid 0 \rangle $  can easily be calculated
using (\ref{61}),(\ref{62})

$$ \langle 0 \mid (\nabla_{\sigma}\bar{\psi}_{\alpha})^d
\bar{\psi}^b_{\beta}G^n_{\mu\nu}\mid 0 \rangle =\frac{g\langle 0
\mid\bar{\psi}\psi\mid 0 \rangle^2}{3^3\cdot2^6}$$

\be
\times [g_{\sigma\mu}\gamma_{\nu}-g_{\sigma\nu}\gamma_{\mu} + i
\varepsilon_{\sigma\mu\nu\lambda}\gamma_5\gamma_{\lambda}]_{\beta\alpha}
(\lambda^n)^{bd}.
\label{63}
\ee
For diagrams in Fig.11 we use the following expansion of the gluon
propagator:

$$S^{np}_{\nu\rho}(x-y,y) = \frac{-i}{(2\pi)^4} g f^{npl}
\int\frac{d^4 k}{k^4} e^{-ik(x-y)}\cdot \left\{\Biggl [
-ik_{\lambda}y_{\alpha}G^l_{\alpha\lambda} -
\frac{2}{3}i(y_{\alpha}y_{\beta}k_{\lambda}\right.$$

$$
-\frac{iy_{\beta}}{k^2}(k^2\delta_{\alpha\lambda}
-2k_{\alpha}k_{\lambda}))(D_{\alpha}G_{\beta\lambda})^l + \frac{1}{3}\Biggl
( y_{\alpha} + \frac{2ik_{\alpha}}{k^2}\Biggr )$$

\be
\times \left.(D_{\lambda}G_{\alpha\lambda})^l\Biggr]\delta_{\nu\rho} + 2
\Biggl [ G^l_{\nu\rho} + 2i \frac{k_{\alpha}}{k^2}
(D_{\alpha}G_{\nu\rho})^l\Biggl ]\right\}
\label{64}
\ee
This expression can
be found by the method of the calculation of the gluon propagator in the
external vacuum gluon field suggested in \cite{15} (see also \cite{16,20}).
The total number of $d=6$ diagrams is enormous -- about 500. Collecting the
results together we get finally the following sum rules for $u^L_{\rho}$

$$xu^L_{\rho}(x) = \frac{3}{4\pi^2}
M^2\frac{g^2_{\rho}}{m^2_{\rho}}e^{m^2_{\rho}/M^2} x^2(1-x)\Biggl [\Biggl
(1+ \Biggl (\frac{a_s(M^2)\cdot ln(Q^2_0/M^2)}{3\pi}\Biggl )$$

\be
\times \Biggl (\frac{1+4x ln(1-x)}{x} - \frac{2(1-2x)ln x}{1-x}\Biggr )
\Biggr )(1-e^{-s_0/M^2})
-\frac{\alpha_s(M^2)\cdot \alpha_s a^2}{\pi^2 \cdot 3^7\cdot 2^6\cdot
M^6}\cdot \frac{\omega(x)}{x^3(1-x)^3}\Biggr ],
\label{65}
\ee
where $a$  and $\omega(x)$  are given by (\ref{29}), (\ref{30}).
A remark here is in order.
In (\ref{65}) (as well in (\ref{28}))
 the
strong coupling constant $g_s$ is  generated in the diagrams in two ways:\\
1) due to explicit quark-gluon interaction  (vertices of a hard gluon
line in the diagrams in Figs.10 and 11, or vertices of external gluon in the
diagrams in Figs.8 and 9), and it is reasonable to take it at the
renormalization point $\mu^2=M^2$;
2) due to the equation of motion, and its normalization point should be
taken in such a way that the quantity $\alpha_s\langle 0 \mid \bar{\psi}\psi
\mid 0 \rangle^2 $ is a renormalization group invariant. The notations in
(\ref{28}),(\ref{65}) reflect this fact.

Sum rules for $u^L_{\rho}(x)$  are
fullfilled in wide region of $x$:  $0.1 < x < 0.85$. Borel mass $M^2$
dependence of $xu^L_{\rho}(x)$ at various $x$ is plotted in Fig.12. One can
see that it is weak in the whole range of $x$, except $x\leq 0.1$ and  $x
\geq 0.75$. As discussed in the Introduction the reason of a more strong
$M^2$-dependence at small and large $x$  is connected with the fact, that
our approach is invalid at small $x$ and $x$, close to 1.  It is manifested
by the blow up of dimension 6 correction at $x\to 0$ and $x\to 1$  in
eq.(\ref{65}). So, the applicability domain of the sum rule can be found
from the sum rule itself. Fig.13 presents $xu^L_{\rho}(x)$  as a function of
$x$. $M^2=1$ GeV$^2$   and $s_0=1.5$ GeV$^2$, $Q_0^2=4$ GeV$^2$  were
chosen, the other set of parameters -- $\Lambda^{LO}_{QCD}$ and $\alpha_s
a^2$  is the same as in the calculation of $xu_{\pi}(x)$.

Valence $u$-quark distribution in transversally polarized $\rho$-meson is
given by

$$
xu^T_{\rho}(x) = \frac{3}{8 \pi^2} g^2_{\rho} e^{m^2_{\rho}/M^2}~
\frac{M^4}{m^4_{\rho}} x \left \{\varphi_0(x) E_1 \Biggl (\frac{s_0}{M^2}
\Biggr ) \Biggl [ 1 +
\frac{1}{3 \pi} ln \Biggl (\frac{Q^2_0}{M^2} \Biggr ) \alpha_s (M^2) \Biggl
( \frac{(4x-1)}{\varphi_0(x)} + \right.  $$

$$ + 4 ln (1-x) - \frac{2(1 - 2x
+ 4x^2) ln x}{\varphi_0(x)} \Biggr ) \Biggr ]- \frac{\pi^2}{6}~
\frac{\langle 0 \vert(\alpha_s/\pi) G^2 \vert 0 \rangle} {M^4 x^2}  $$

$$
+\frac{1}{2^8 \cdot 3^5 M^6 x^3 (1-x)^3} \langle 0 \vert
g^3 f^{abc} G^a_{\mu \nu} G^b_{\nu \lambda} G^c_{\lambda \mu} \vert 0
\rangle \xi (x)
$$

\be
\left. +\frac{\alpha_s(M^2) (\alpha_s a^2)}{2^5 \cdot 3^8 \pi^2 M^6 x^3 (1
- x)^3 }\chi (x) \right\}
\label{66}
\ee

 \be
 \xi(x) = -1639 + 8039 x - 15233 x^2 +
10055 x^3 - 624x^4 - 974 x^5
\label{67}
\ee

$$
\chi(x) = 8513 - 41692 x + 64589 x^2 - 60154 x^3 + 99948 x^4
$$
$$
- 112516 x^5 + 45792 x^6 + (- 180 - 8604 x + 53532 x^2
$$
\be
- 75492 x^3 - 28872x^4 + 109296 x^5 - 55440 x^6) ln 2
\label{68}
\ee
The standard value \cite{14} of gluonic  condensate $\langle 0 \mid
(\alpha_s/\pi)G^2\mid 0 \rangle=0.012~GeV^4$ was taken in numerical
calculations. Unlike the cases of $u_{\pi}(x)$ and $u^L_{\rho}(x)$,
dimension-6 power correction, proportional to $\langle 0 \vert g f^{abc}
G^a_{\mu \nu} G^b _{\nu \lambda} G^c _{\lambda \mu}\mid 0 \rangle$, is not
cancelled here.  Its contribution is calculated with the help of the
instanton gas model -- eq.(\ref{60}).
The effective instanton radius $\rho_c$ was chosen as
$\rho_c = 0.5 fm$. This value is between the estimations of \cite{18}
($\rho_c = 1 fm$) and \cite{19} ($\rho_c = 0.33 fm$). (In the recent paper
\cite{21} the arguments were presented that the liquid gas instanton model
overestimates higher order gluonic condensates and in order to correct this
effect larger values of $\rho_c$ comparing with \cite{19} should be used).
Borel mass dependence of $xu_T(x)$ is
shown in Fig.14.
As is seen from Fig.14, in the interval $0.2 < x < 0.65$ the $M^2$-dependence
is weak at $0.8 < M^2 < 1.2 ~GeV^2$. Fig.15 shows $xu_T(x)$ at $M^2 = 1
~GeV^2$ and $Q^2_0 = 4 ~GeV^2$.  Dashed lines demonstrate the influence of
the variation of $\rho_c$ in the final result:  the lower line corresponds
to $\rho_c = 0.6 fm$ and the upper -- to $\rho_c = 0.4 fm$.  Our results are
reliable at $0.2 < x < 0.65$, where $d = 4$ and $d = 6$ (separately) power
corrections comprise less than 30\% of the bare loop contribution. ($\langle
0 \vert (\alpha_s/\pi) G^2 \vert 0 \rangle$ and $\langle 0 \vert g^3 f^{abc}
G^a_{\mu\nu}G^b_{\nu\lambda}G^c_{\lambda\mu} \vert 0 \rangle$ contributions
are of the opposite sign and compensate one another, $\alpha_s(M^2)\alpha_s
a^2$ contribution is negligible.) At $\rho_c = 0.4 fm$ the applicability
domain shrinks to $0.25 < x < 0.6$. The moments of quark distributions in
longitudinal $\rho$-meson are calculated in the same way, as it was done in
the case of pion: by matching with Regge behaviour $u(x) \sim 1/\sqrt{x}$ at
low $x$ and with quark counting rule $u(x) \sim (1 - x)^2$ at large $x$. The
matching points were chosen as $x = 0.10$ at low $x$ and $x = 0.80$ at large
$x$. The numerical values of moments for longitudinally polarized $\rho$ are

$$
{\cal{M}}^L_1 = \int \limits ^{1}_{0} dx u^L_{\rho} (x) = 1.06 ~~~ (1.05)
$$
\be
{\cal{M}}^L_2= \int \limits ^{1}_{0} x dx u^L_{\rho} (x) = 0.39 ~~~ (0.37)
\label{69}
\ee
The values of momenta, obtained by assuming that $u(x) \sim (1 - x)$ at
large $x$ are given in the parenthesis.

Reliable calculation of moments in the case of transversally polarized
$\rho$-meson is impossible, because of a narrow applicability domain in $x$
and the form of $u$-quark distribution -- Fig.15, which does not allow soft
matching with expected behaviour $xu^T_{\rho}(x)$ at small and large $x$.

The comparison of $u$-quark distributions in longitudinally and
transversally polarized $\rho$-mesons shows a strong difference of them: the
curvatures have the opposite sign: $u^L_{\rho}(x)$  has a maximum at
intermediate $x\sim 0.5$ while $u^T_{\rho}(x)$ has a minimum there.
Strongly different are also the second moments
in pion and in longitudinal $\rho$-meson: the total part of the momentum,
carried by valence quarks and antiquarks $-(u+\bar{d})$   in longitudinal
$\rho$-meson is about 0.8, while in pion it is much less -- about 0.4-0.5.
The same amount as in pion one may expect in transverse $\rho$ (Fig.15),
nevertheless, that it is impossible to calculate a precise number.
Therefore, pion and transverse $\rho$-mesons in this aspect behave like a
nucleon, where about 50\% of total momentum is carried by gluons and sea
quarks. In longitudinal $\rho$ the situation is different -- only about 20\%
of momentum is left for gluons and sea quarks. It must be mentioned, that in
case of transverse $\rho$ the accuracy of our results are worse than for
longitudinal $\rho$, because the contribution of higher order terms of OPE
is larger and the applicability domain in $x$  is narrower. This fact,
however, does not change the qualitative conclusion formulated above.
Now let us discuss the nonpolarizied $\rho$-meson case.
Quark distribution function $u(x)$ in this case is equal to:

$$u_{\rho}(x)= (u^L_{\rho}(x) + 2 u^T_{\rho}(x))/3$$
and we can determine $u(x)$ only in the region, where sum rules  for
$u^L_{\rho}(x)$ and $u^T_{\rho}$
are fulfiled, i.e. $0.2 \la x\leq 0.65$.
In this region $u_{\rho}(x)$ is found to be very close to $u_{\pi}(x)$
(the difference  in whole range of $x$ is not  more than 10-15\%).

\section{Summary and discussion}

\hspace{5mm} Let us first discuss the accuracy of our results. In case of
$u$-quark distribution in  pion the main uncertainty comes from the
magnitude of $\alpha_s \langle 0 \mid \bar{\psi}\psi \mid 0 \rangle^2$. For
renorminvariant quantity $(2\pi)^4
\alpha_s \langle \mid \bar{\psi}\psi \mid 0 \rangle^2$  in our calculations
we took the value 0.13 Gev$^6$. In fact, however, it is uncertain by a
factor of 2. (Recent determination \cite{22} of this quantity from
$\tau$-decay data indicates that it may be two times larger). Also the
perturbative corrections introduce some uncertainties, especially at large
$x,(x > 0.6)$ where the accounted LO correction is large. (E.g. instead of
$\Lambda_{QCD}=200$ MeV the value $\Lambda_{QCD}=250$ MeV could be taken).
The estimation of both effects shows, that they may result in 10-15\%
variation of $xu_{\pi}(x)$ -- increasing at $x < 0.3$ and decreasing at $x >
0.3$. 
One should note, that our estimation of the second moment of quark distribution 
in $\pi$-meson (32)  differ from those obtained in 
\cite{23}. The origin of this discrepancy is not completely clear now:
partly it could be related with the difference of the value of parameters 
used in \cite{23}  ($\Lambda^{LO}_{QCD}=100 MeV,~ Q^2_0=1 GeV^2$)
from our choise of parameters, and partly, 
maybe, with the uncertainty  of our estimations of the moments. 

In case of $u$-quark distribution in longitudinally polarized $\rho$-meson
the uncertainties in $\alpha_s\langle 0\mid \bar{\psi}\psi \mid 0 \rangle^2$
do not play any role, because of higher values of $M^2$  and the main source
of them is the perturbative corrections. They influence only high $x$
domain, $x > 0.5$.

The accuracy of our results for $u$-quark distribution in transversally
polarized $\rho$-meson is worse, because of a large role of $d=4$  and $d=6$
gluonic condensate contributions. The variation of $xu^T_{\rho}(x)$ arising
from uncertainties of $d=6$ gluonic condensates was shown in Fig.15. The
gluonic condensate $\langle 0 \mid(\alpha_s/\pi)G^2 \mid 0 \rangle$  is also
uncertain by a factor 1.5. It may result in 30-40\% variation of
$xu_{\rho}^T$  at $x\approx 0.3-0.4$, but much less at $x\approx
0.5-0.6$. At least, these uncertainties do not influence the shape of
$u$-quark distribution. (The LO perturbative corrections are
no more than 20\% at small $x$ and negligible at large $x$).

Fig.16 gives the comparison of valence $u$-quark distributions in pion,
longitudinally and transversally polarized $\rho$-mesons. The shapes of
curves  are quite different, especially of $xu^T_{\rho}(x)$  in comparison
with $xu^L_{\rho}(x)$ and $xu_{\pi}(x)$. Any uncertainties in our
calculations cannot influence this basic conclusion. The values of moments
of quark distributions are also different -- see the discussion at the end
of Sec.4.

The main physical conclusion is: the quark distributions in pion and
$\rho$-meson have not to much in common. The specific properties of pion, as
a Goldstone boson manifest themselves  in different quark distributions in
comparison with $\rho$.   $SU(6)$  symmetry, probably, may take place for
static properties of $\pi$ and $\rho$, but not for their internal structure.
This fact is not surprising. Quark distributions have sense in fastly moving
hadrons. However, $SU(6)$ symmetry cannot be selfconsistently generalized to
relativistic case \cite{26}. We have no explanation, why $u$-quark
distributions in pion and nonpolarized $\rho$-meson at $0.2 < x < 0.65$ are
close to one another -- is it a pure accident or there are some deep reasons
for it.

\vspace{7mm}

{\bf \Large Acknowledgements}

\bigskip

\noindent
The research described in this publication was made possible  in part
by Award No RP2-2247 of U.S.  Civilian Research and Development Foundation
for Independent  States of Former Soviet Union (CRDF) and by Russian Found
of Basic Research grant 00-02-17808.

\newpage

\begin{figure}
\epsfxsize=10cm
\epsfbox{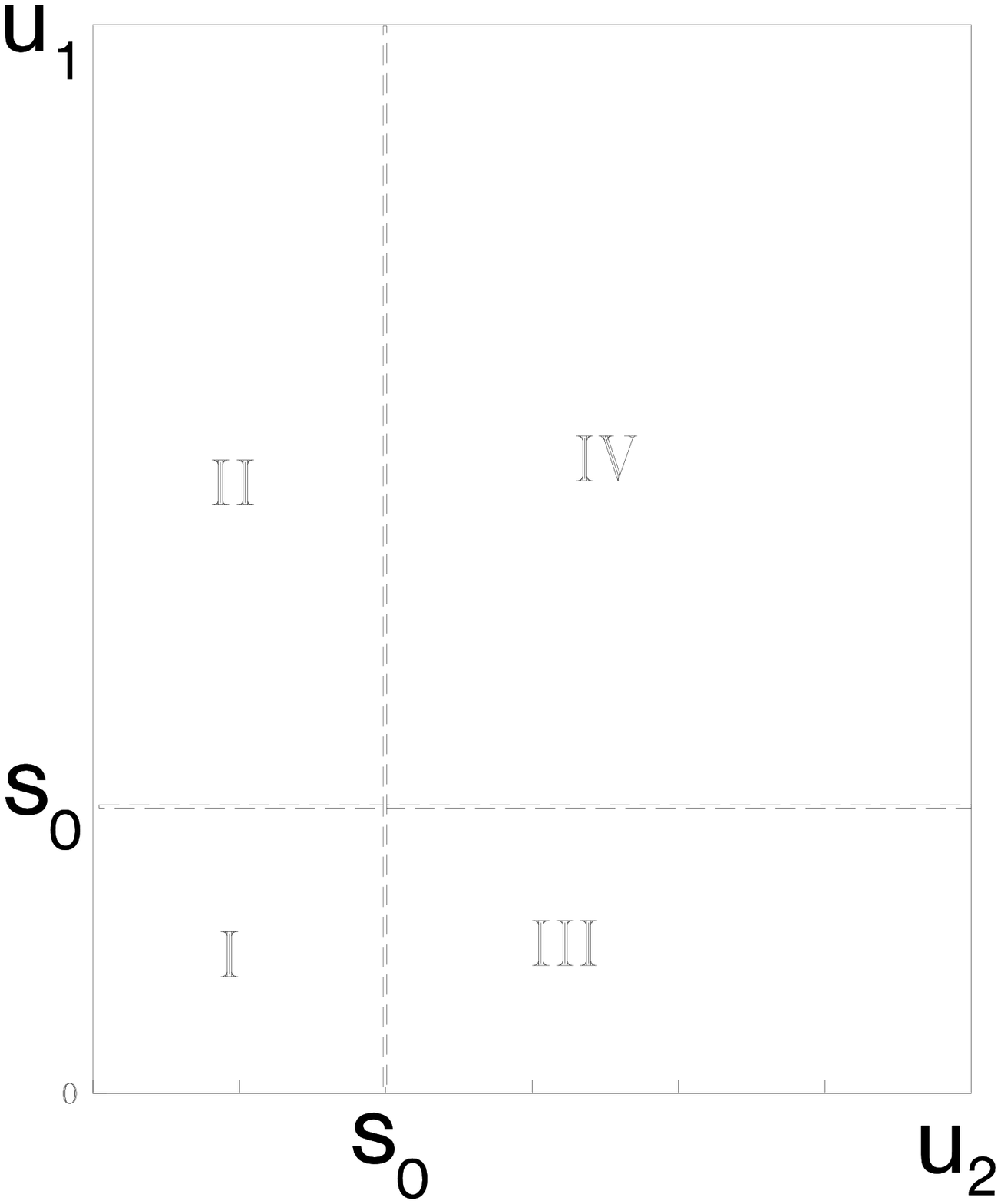}
\caption{ Integration region in the double dispersion representation.}
\end{figure}
\newpage

\begin{figure}
\epsfxsize=10cm
\epsfbox{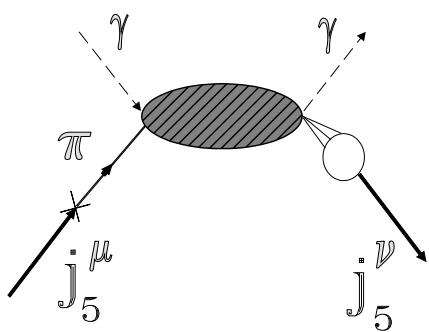}
\caption{ Example of the non-diagonal transition}
\end{figure}
\newpage

\begin{figure}
\epsfxsize=10cm
\epsfbox{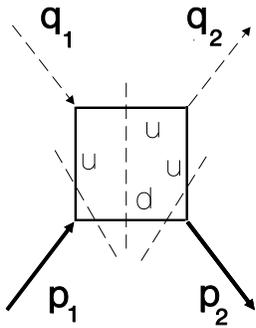}
\caption{ Diagram, corresponding to the unit operator contribution. Dashed
lines with arrows correspond to the photon, thick solid - to hadron current}
\end{figure}
\newpage

\begin{figure}
\epsfxsize=10cm
\epsfbox{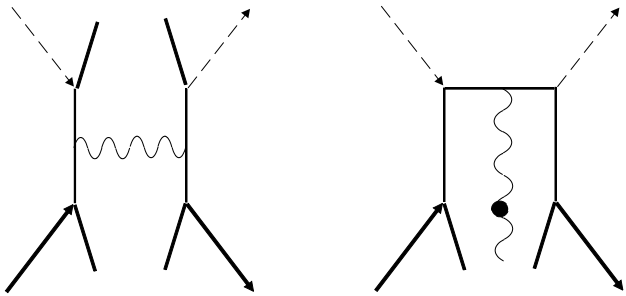}
\caption{ Examples of the non-loop diagrams of dimension 4.
Wave lines correspond to gluons, dot means derivative, other
notations as in Fig.3.}
\end{figure}
\newpage

\begin{figure}
\epsfxsize=10cm
\epsfbox{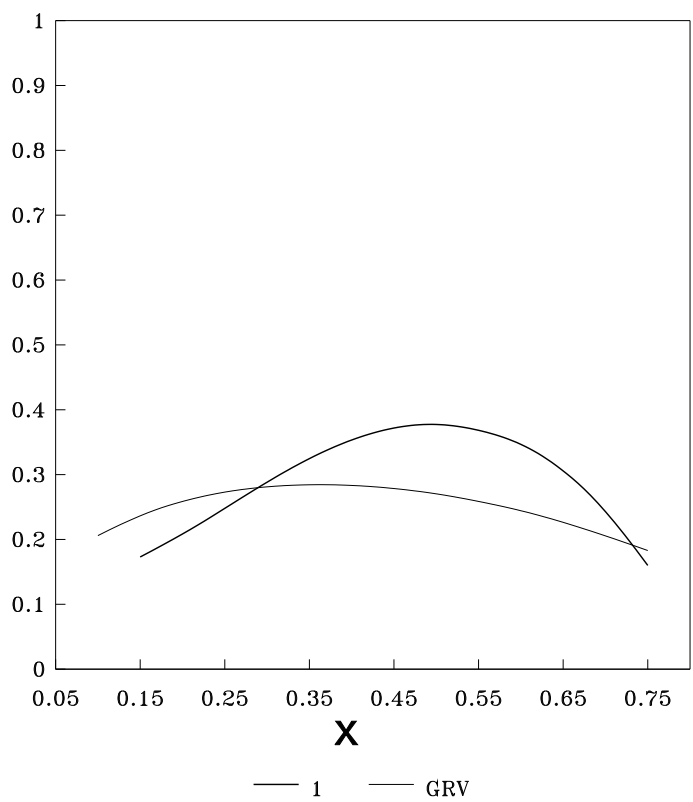}
\caption{ Quark distribution function in pion, noted "1". For comparison
 the fit from [6], labelled "GRV", is shown. }
\end{figure}
\newpage

\begin{figure}
\epsfxsize=10cm
\epsfbox{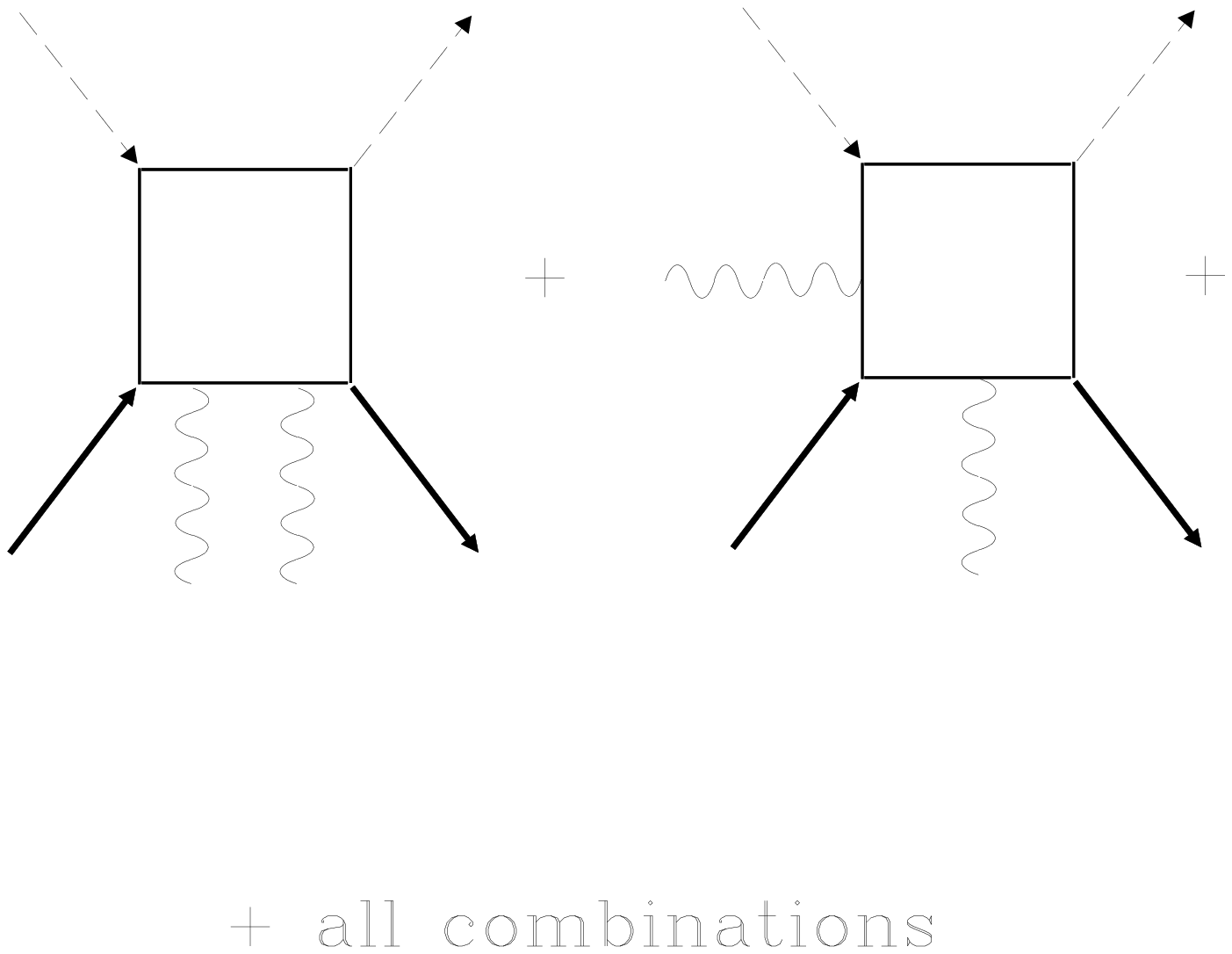}
\caption{ Diagrams, corresponding to the $d=4$ operator contribution.
Dashed lines with arrows correspond to the photon, thick solid - to 
the hadron current, wave lines correspond to the external gluon field}
\end{figure}
\newpage

\begin{figure}
\epsfxsize=10cm
\epsfbox{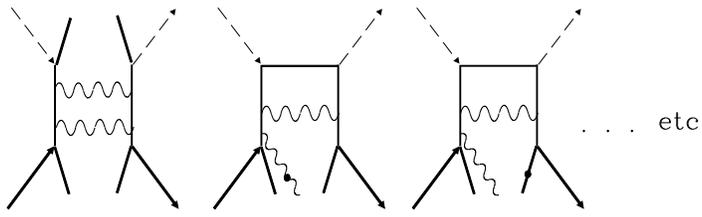}
\caption{ Dimension 6 diagrams corresponding to evolution of non-loop diagrams.
All notations as in Fig.4}

\end{figure}

\newpage

\begin{figure}
\epsfxsize=10cm
\epsfbox{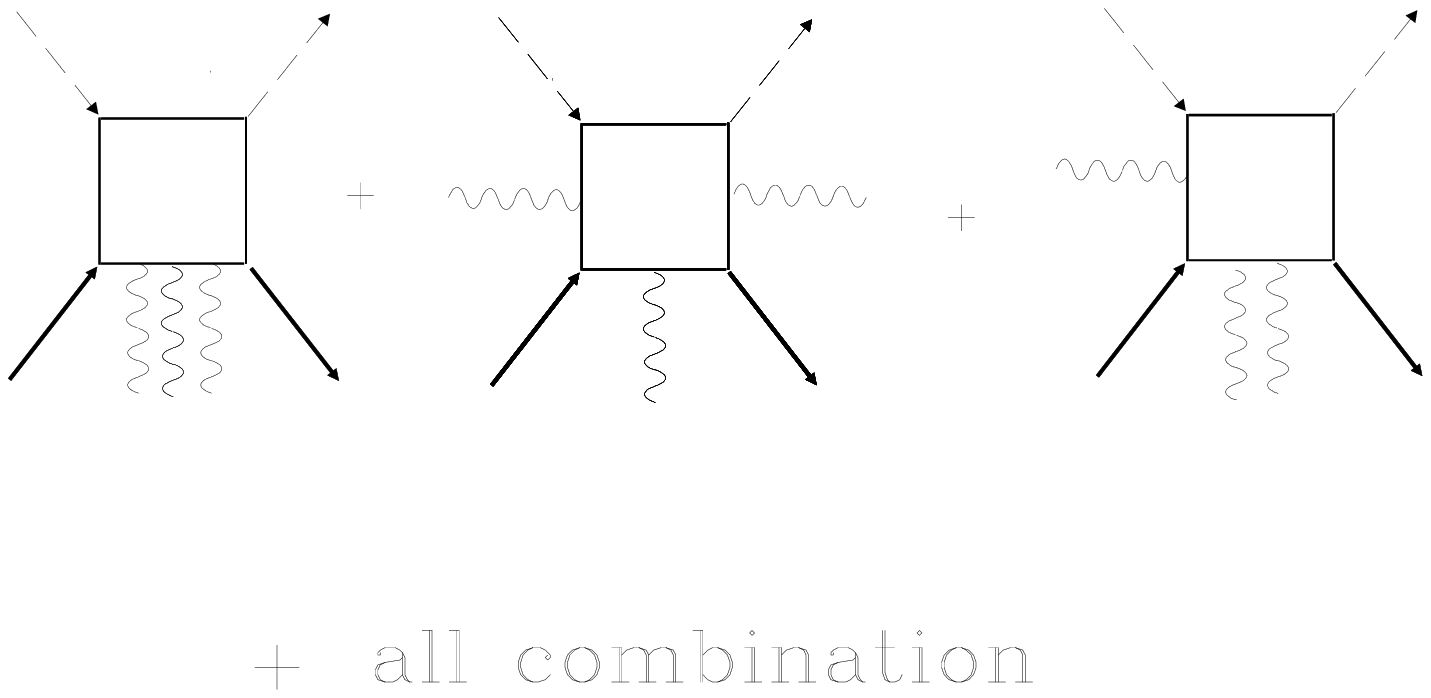}
\caption{ Diagrams of dimension 6, see text. All notations as in Fig.4}
\end{figure}
\newpage

\begin{figure}
\epsfxsize=10cm
\epsfbox{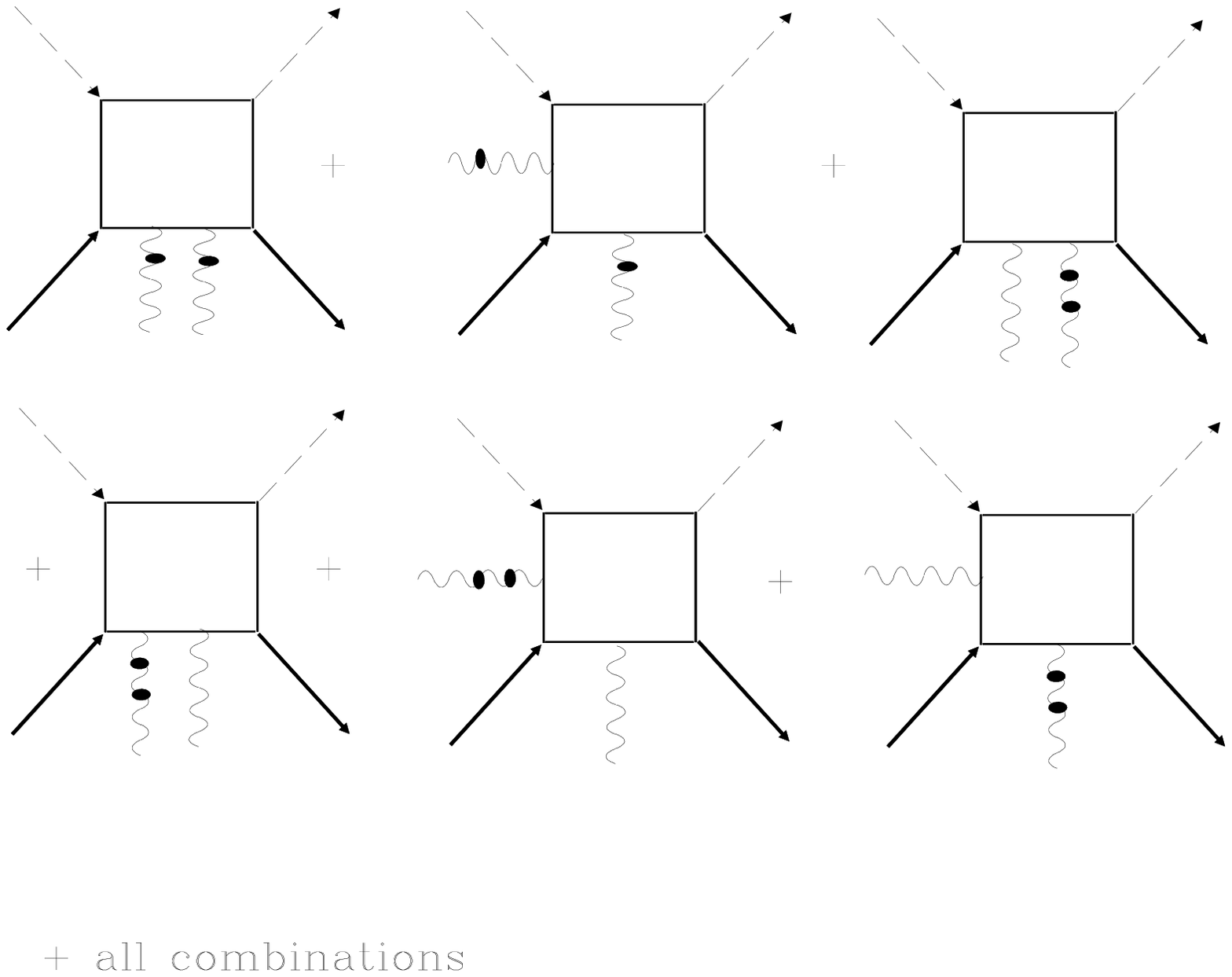}
\caption{ Diagrams of dimension 6. External lines with
dots correspond to derivatives in external fields.
All notations as in Fig.6}
\end{figure}
\newpage

\begin{figure}
\epsfxsize=10cm
\epsfbox{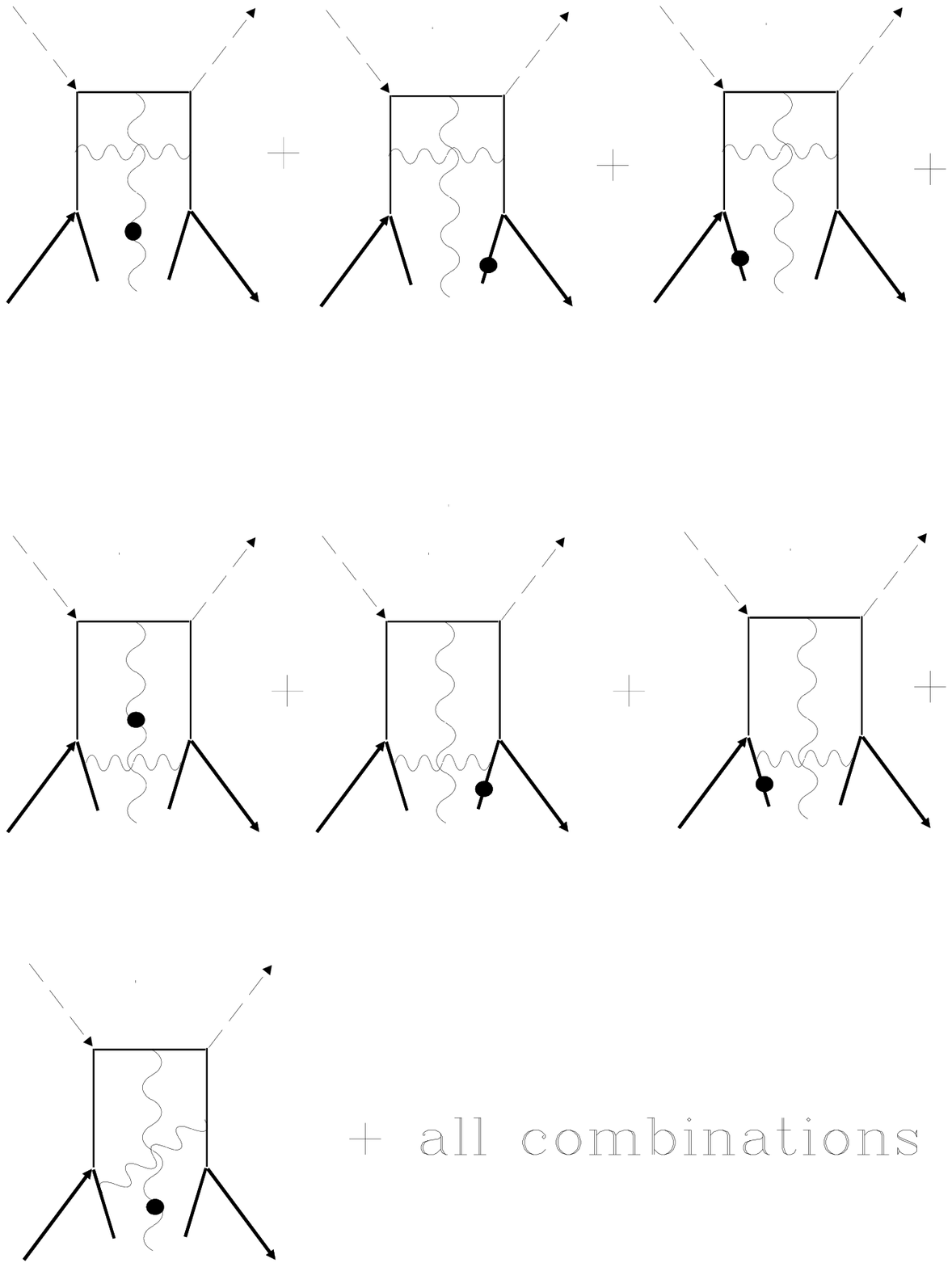}
\caption{Diagrams of dimension 6, corresponding to the quark propagator
expansion. All notations as in Fig.4}
\end{figure}
\newpage

\begin{figure}
\epsfxsize=10cm
\epsfbox{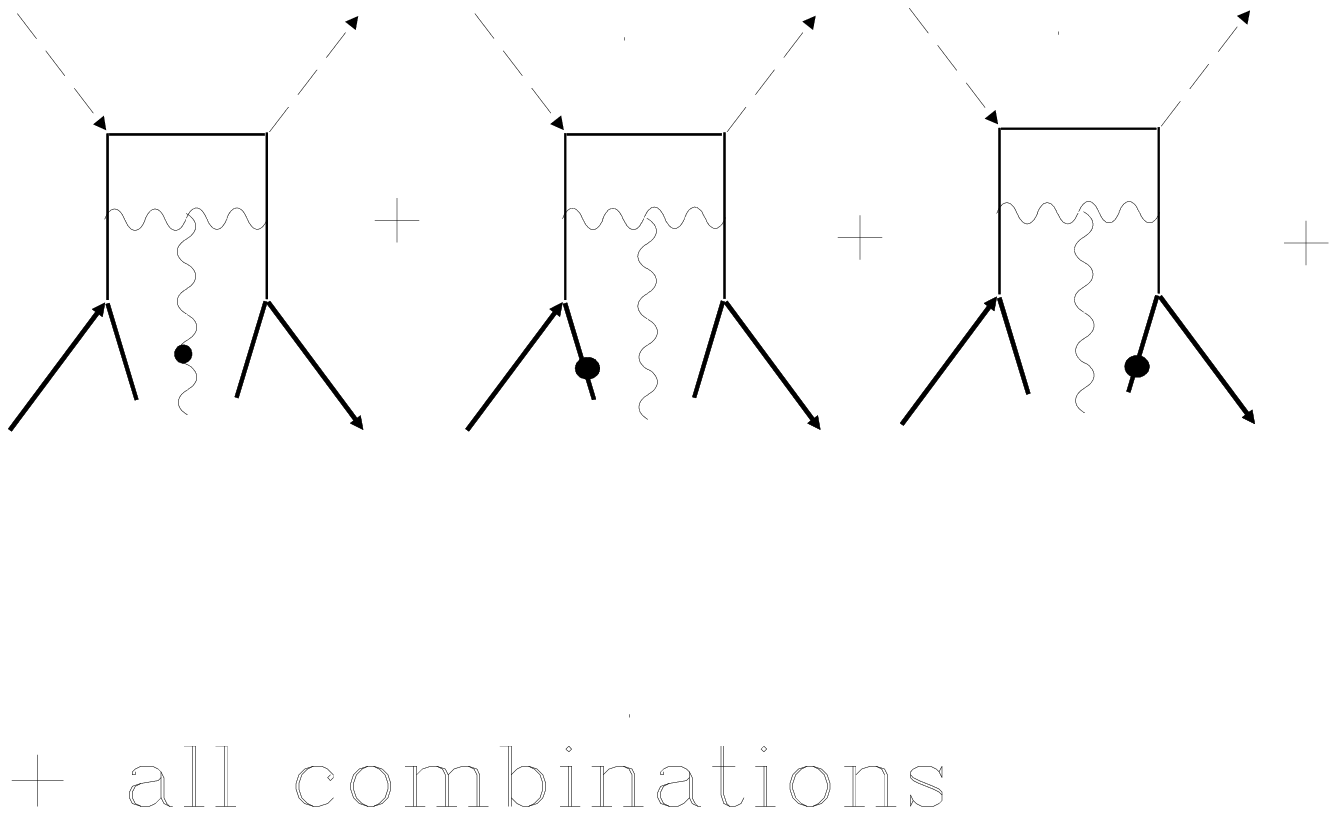}
\caption{Diagrams of dimension 6, corresponding to the quark and gluon
propagator expansion. All notations as in Fig.4}
\end{figure}
\newpage

\begin{figure}
\epsfxsize=10cm
\epsfbox{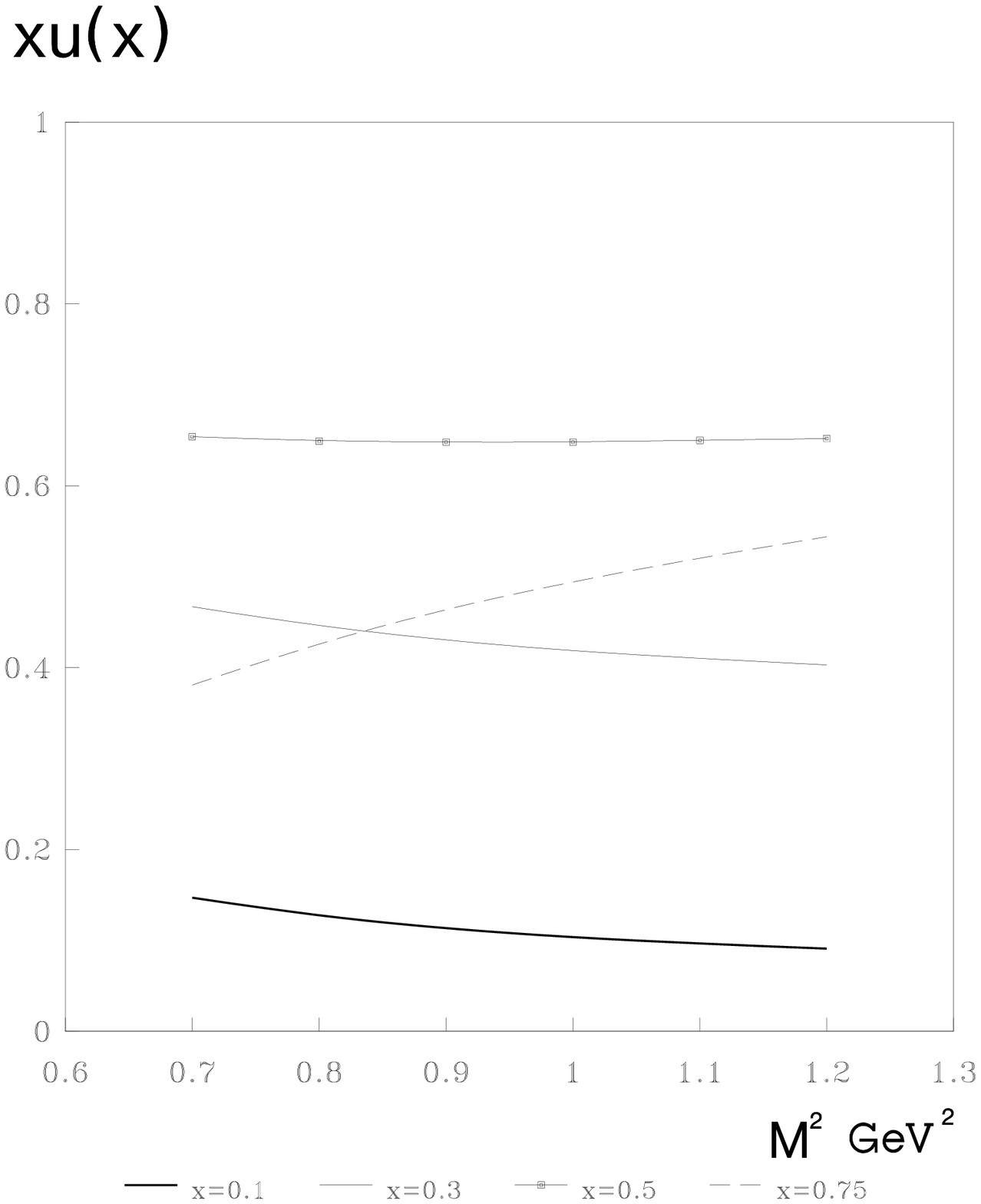}
\caption{Borel mass dependence of the quark distribution function $xu^L_{\rho}(x)$  at various $x$}
\end{figure}
\newpage

\begin{figure}
\epsfxsize=10cm
\epsfbox{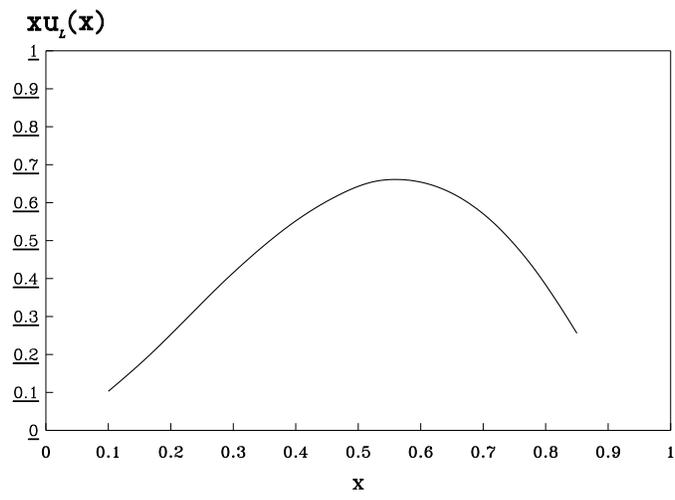}
\caption{ quark distribution function $xu^L_{\rho}(x)$ }
\end{figure}
\newpage

\begin{figure}
\epsfxsize=10cm
\epsfbox{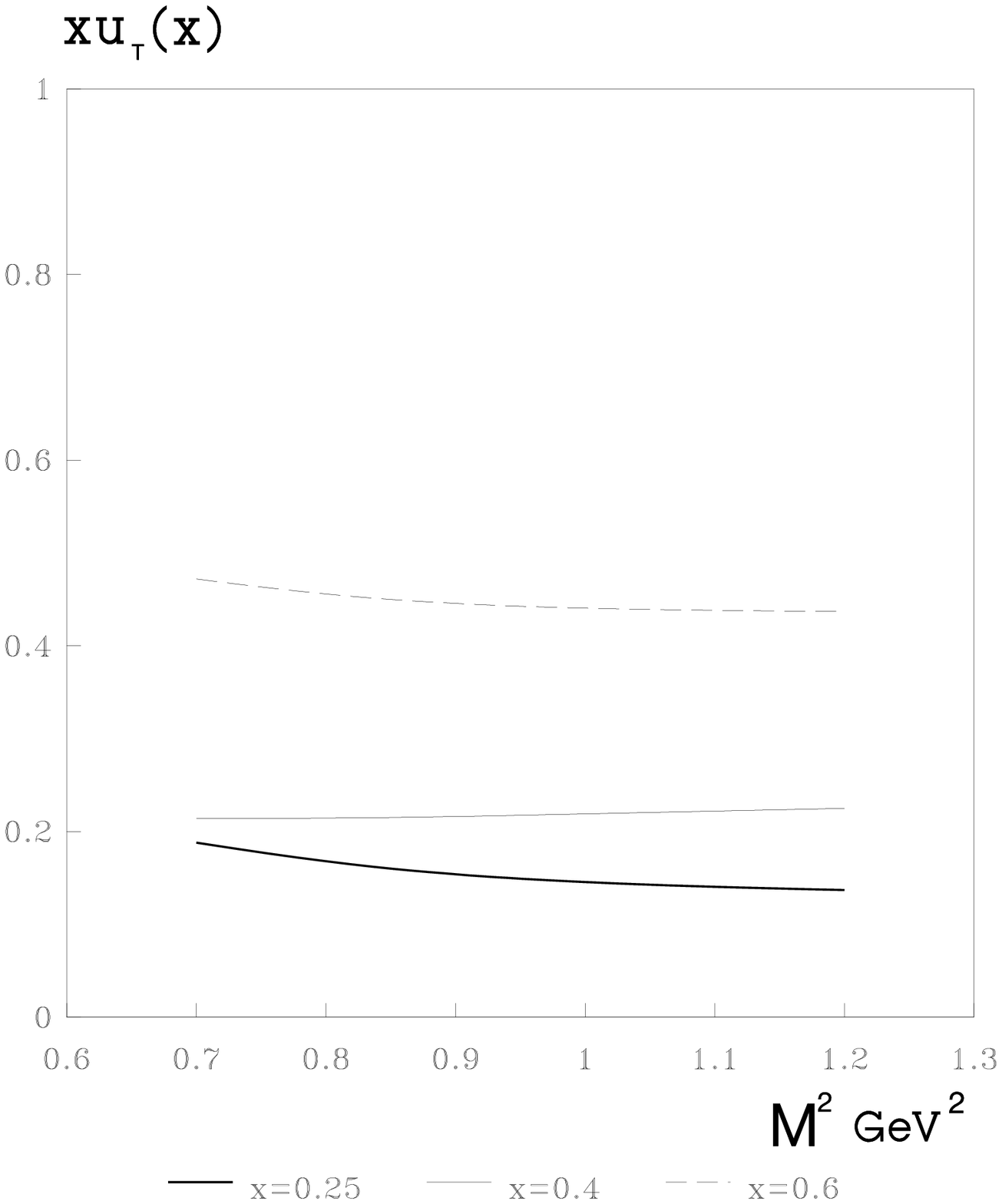}
\caption{ Borel mass dependence of the quark distribution function  $xu^T_{\rho}(x)$ at various $x$}
\end{figure}
\newpage

\begin{figure}
\epsfxsize=10cm
\epsfbox{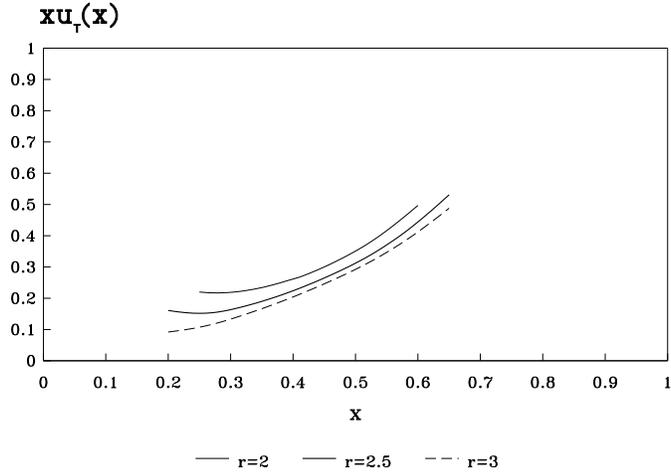}
\caption{$xu^T(x)$ for transversally polarizied $\rho$-meson at three
choices of instanton radius $\rho=2, 2.5, 3~ GeV^{-1}$ (correspondingly,
curves are labelled by  $r=2, r=2.5, r=3$)}
\end{figure}

\newpage

\begin{figure}
\epsfxsize=10cm
\epsfbox{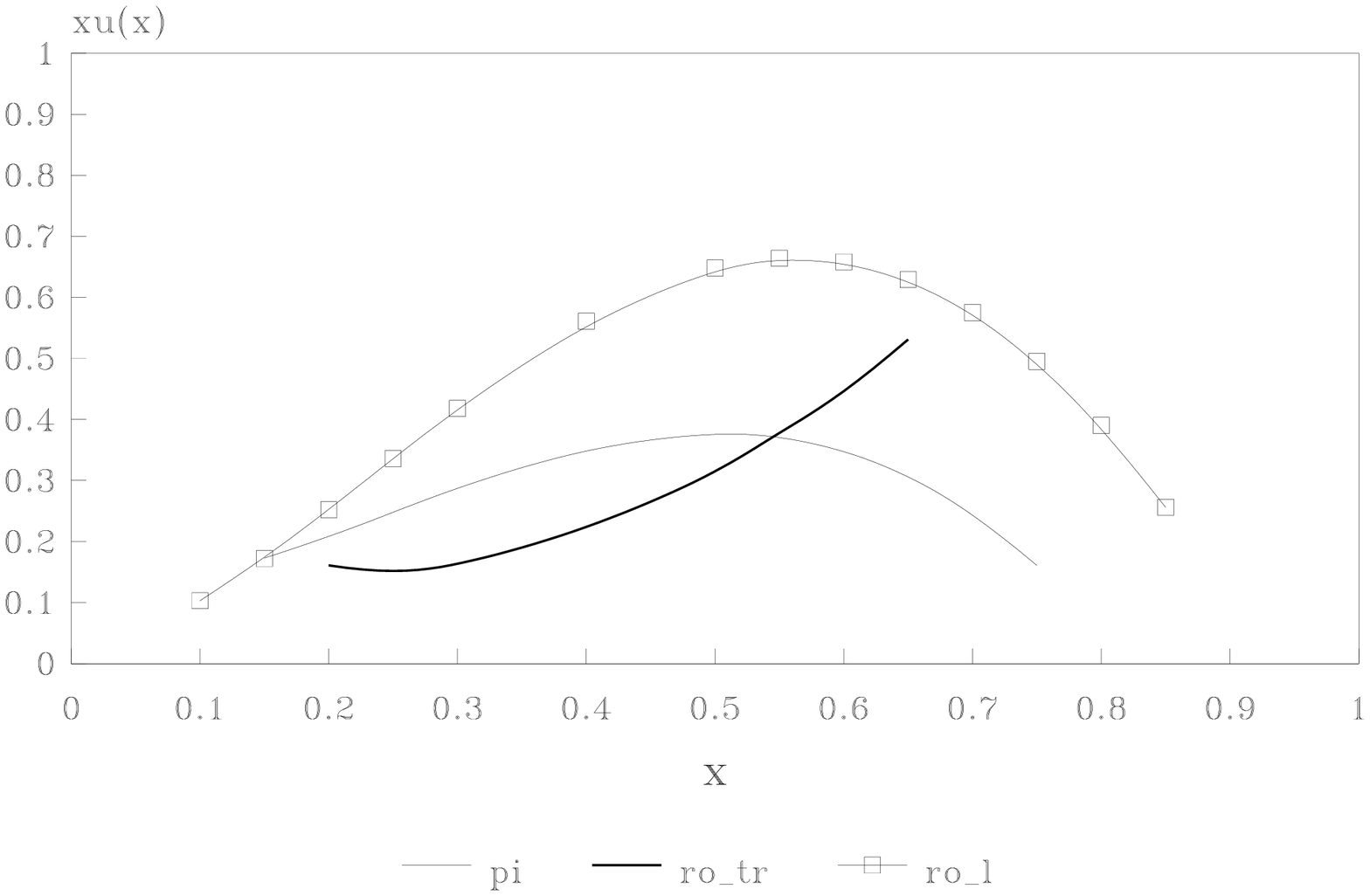}
\caption{xu(x) for $\rho^T$-(curve is labelled by ro-tr), $\rho^L$ (is labelled by ro-l) and $\pi$-meson (pi)}
\end{figure}
\newpage

\end{document}